
\input harvmac.tex
\input epsf.tex

\def\figin{\epsfcheck\figin}\def\figins{\epsfcheck\figins}
\def\epsfcheck{\ifx\epsfbox\UnDeFiNeD
\message{(NO epsf.tex, FIGURES WILL BE IGNORED)}
\gdef\figin##1{\vskip2in}\gdef\figins##1{\hskip.5in}
\else\message{(FIGURES WILL BE INCLUDED)}%
\gdef\figin##1{##1}\gdef\figins##1{##1}\fi}
\def\DefWarn#1{}
\def\figinsert{\goodbreak\midinsert}
\def\ifig#1#2#3{\DefWarn#1\xdef#1{fig.~\the\figno}
\writedef{#1\leftbracket fig.\noexpand~\the\figno} %
\figinsert\figin{\centerline{#3}}\medskip\centerline{\vbox{\baselineskip12pt
\advance\hsize by -1truein\noindent\footnotefont{\bf
Fig.~\the\figno:} #2}}
\bigskip\endinsert\global\advance\figno by1}


\def\la{ \langle}
\def\ra {\rangle}
\def \la {\langle}
\def \ra {\rangle}
\def \pa {\partial}

\def \eps {\epsilon}


\def\bos{{\rm bos}}
\def\fer{{\rm fer}}
\def\odd{ {\rm odd}}
\def\tN{{\tilde N }}
\def\eff{{ \rm eff}}
\def\tla{{\tilde \lambda}}




\lref\HertogDR{
  T.~Hertog and K.~Maeda,
JHEP {\bf 0407}, 051 (2004).
[hep-th/0404261].
}

\lref\HertogHU{
  T.~Hertog and G.~T.~Horowitz,
JHEP {\bf 0504}, 005 (2005).
[hep-th/0503071].
}

\lref\WittenUA{
  E.~Witten,
[hep-th/0112258].
}


\lref\VasilievDN{
  M.~A.~Vasiliev,
Int.\ J.\ Mod.\ Phys.\ D {\bf 5}, 763 (1996).
[hep-th/9611024].
}

\lref\VasilievBA{
  M.~A.~Vasiliev,
In *Shifman, M.A. (ed.): The many faces of the superworld* 533-610.
[hep-th/9910096].
}

\lref\SezginEH{
  E.~Sezgin and P.~Sundell,
In *Goeteborg 1998, Novelties in string theory* 241-269.
[hep-th/9903020].
}

\lref\SezginRT{
  E.~Sezgin and P.~Sundell,
Nucl.\ Phys.\ B {\bf 644}, 303 (2002), [Erratum-ibid.\ B {\bf 660}, 403 (2003)].
[hep-th/0205131].
}

\lref\SezginPT{
  E.~Sezgin and P.~Sundell,
JHEP {\bf 0507}, 044 (2005).
[hep-th/0305040].
}

\lref\KlebanovJA{
  I.~R.~Klebanov and A.~M.~Polyakov,
Phys.\ Lett.\ B {\bf 550}, 213 (2002).
[hep-th/0210114].
}

\lref\SundborgWP{
  B.~Sundborg,
Nucl.\ Phys.\ Proc.\ Suppl.\  {\bf 102}, 113 (2001).
[hep-th/0103247].
}

\lref\wittentalk{
E. Witten, Talk at the John Schwarz 60-th Birthday Symposium,
~~~~~~~~~~~ // ~~~~~~~~~~~~~~ ~~~~ ~~~~~ ~~~~  ~~~~
http://theory.caltech.edu/jhs60/witten/1.html.
 }

\lref\ColemanAD{
  S.~R.~Coleman, J.~Mandula,
Phys.\ Rev.\  {\bf 159}, 1251-1256 (1967).
}

\lref\HaagQH{
  R.~Haag, J.~T.~Lopuszanski, M.~Sohnius,
Nucl.\ Phys.\  {\bf B88}, 257 (1975).
}

\lref\HeidenreichXI{
  W.~Heidenreich,
J.\ Math.\ Phys.\  {\bf 22}, 1566 (1981).
}

\lref\SagnottiQP{
  A.~Sagnotti,
[arXiv:1112.4285 [hep-th]].
}

\lref\AnselmiBH{
  D.~Anselmi,
Nucl.\ Phys.\ B {\bf 541}, 323 (1999).
[hep-th/9808004].
}

\lref\AnselmiMS{
  D.~Anselmi,
Nucl.\ Phys.\ B {\bf 541}, 369 (1999).
[hep-th/9809192].
}

\lref\RuhlBW{
  W.~Ruhl,
[hep-th/0607197].
}

\lref\GrossUE{
  D.~J.~Gross,
Phys.\ Rev.\ Lett.\  {\bf 60}, 1229 (1988)..
}

\lref\MooreQE{
  G.~W.~Moore,
In *Boston 1993, Proceedings, Supersymmetry and unification of fundamental interactions, SUSY 93* 540-550.
[hep-th/9308052].
}

\lref\HennMW{
  J.~Henn, C.~Jarczak and E.~Sokatchev,
Nucl.\ Phys.\ B {\bf 730}, 191 (2005).
[hep-th/0507241].
}

\lref\BianchiWX{
  M.~Bianchi, J.~F.~Morales and H.~Samtleben,
JHEP {\bf 0307}, 062 (2003).
[hep-th/0305052].
}

\lref\MaldacenaNZ{
  J.~M.~Maldacena and G.~L.~Pimentel,
JHEP {\bf 1109}, 045 (2011).
[arXiv:1104.2846 [hep-th]].
}

\lref\SachdevPR{
  S.~Sachdev,
Phys.\ Lett.\ B {\bf 309}, 285 (1993).
[hep-th/9305131].
}

\lref\GrossJV{
  D.~J.~Gross and A.~Neveu,
Phys.\ Rev.\ D {\bf 10}, 3235 (1974)..
}

\lref\GiombiKC{
  S.~Giombi, S.~Minwalla, S.~Prakash, S.~P.~Trivedi, S.~R.~Wadia and X.~Yin,
[arXiv:1110.4386 [hep-th]].
}

\lref\GirardelloPP{
  L.~Girardello, M.~Porrati and A.~Zaffaroni,
Phys.\ Lett.\ B {\bf 561}, 289 (2003).
[hep-th/0212181].
}

\lref\MaldacenaJN{
  J.~Maldacena and A.~Zhiboedov,
[arXiv:1112.1016 [hep-th]].
}

\lref\AharonyJZ{
  O.~Aharony, G.~Gur-Ari and R.~Yacoby,
[arXiv:1110.4382 [hep-th]].
}

\lref\GiombiYA{
  S.~Giombi and X.~Yin,
[arXiv:1105.4011 [hep-th]].
}

\lref\GiombiRZ{
  S.~Giombi, S.~Prakash and X.~Yin,
[arXiv:1104.4317 [hep-th]].
}

\lref\GiombiVG{
  S.~Giombi and X.~Yin,
JHEP {\bf 1104}, 086 (2011).
[arXiv:1004.3736 [hep-th]].
}

\lref\GiombiWH{
  S.~Giombi and X.~Yin,
JHEP {\bf 1009}, 115 (2010).
[arXiv:0912.3462 [hep-th]].
}

\lref\FitzpatrickZM{
  A.~L.~Fitzpatrick, E.~Katz, D.~Poland and D.~Simmons-Duffin,
JHEP\ {\bf 1107}, 023  (2011).
[arXiv:1007.2412 [hep-th]].
}
\lref\BelavinVU{
  A.~A.~Belavin, A.~M.~Polyakov, A.~B.~Zamolodchikov,
Nucl.\ Phys.\  {\bf B241}, 333-380 (1984).
}
\lref\SagnottiAT{
  A.~Sagnotti and M.~Taronna,
Nucl.\ Phys.\ B\ {\bf 842}, 299  (2011).
[arXiv:1006.5242 [hep-th]].
}

\lref\ColemanAD{
  S.~R.~Coleman, J.~Mandula,
Phys.\ Rev.\  {\bf 159}, 1251-1256 (1967).
}

\lref\ZamolodchikovXM{
  A.~B.~Zamolodchikov, A.~B.~Zamolodchikov,
Annals Phys.\  {\bf 120}, 253-291 (1979).
}
\lref\ShenkerZF{
  S.~H.~Shenker and X.~Yin,
[arXiv:1109.3519 [hep-th]].
}
\lref\KlebanovJA{
  I.~R.~Klebanov, A.~M.~Polyakov,
Phys.\ Lett.\  {\bf B550}, 213-219 (2002).
[hep-th/0210114].
}
\lref\FradkinQY{
  E.~S.~Fradkin and M.~A.~Vasiliev,
  Nucl.\ Phys.\  B {\bf 291}, 141 (1987).
}
\lref\GiombiRZ{
  S.~Giombi, S.~Prakash, X.~Yin,
[arXiv:1104.4317 [hep-th]].
}

\lref\MikhailovBP{
  A.~Mikhailov,
[hep-th/0201019].
}

\lref\DouglasRC{
  M.~R.~Douglas, L.~Mazzucato and S.~S.~Razamat,
Phys.\ Rev.\ D\ {\bf 83}, 071701  (2011).
[arXiv:1011.4926 [hep-th]].
}

\lref\CostaMG{
  M.~S.~Costa, J.~Penedones, D.~Poland, S.~Rychkov,
[arXiv:1107.3554 [hep-th]].
}

\lref\AnninosUI{
  D.~Anninos, T.~Hartman, A.~Strominger,
[arXiv:1108.5735 [hep-th]].
}

\lref\HaagQH{
  R.~Haag, J.~T.~Lopuszanski, M.~Sohnius,
Nucl.\ Phys.\  {\bf B88}, 257 (1975).
}

\lref\papadodimas{
  S.~El-Showk, K.~Papadodimas,
[arXiv:1101.4163 [hep-th]].
}

\lref\FateevZH{
  V.~A.~Fateev, S.~L.~Lukyanov,
Int.\ J.\ Mod.\ Phys.\  {\bf A3}, 507 (1988).
}
\lref\HeemskerkPN{
  I.~Heemskerk, J.~Penedones, J.~Polchinski and J.~Sully,
  JHEP {\bf 0910}, 079 (2009)
  [arXiv:0907.0151 [hep-th]].
}
\lref\GiombiWH{
  S.~Giombi, X.~Yin,
JHEP {\bf 1009}, 115 (2010).
[arXiv:0912.3462 [hep-th]].
}

\lref\MaldacenaNZ{
  J.~M.~Maldacena, G.~L.~Pimentel,
JHEP {\bf 1109}, 045 (2011).
[arXiv:1104.2846 [hep-th]].
}

\lref\GiombiKC{
  S.~Giombi, S.~Minwalla, S.~Prakash, S.~P.~Trivedi, S.~R.~Wadia, X.~Yin,
[arXiv:1110.4386 [hep-th]].
}

\lref\AharonyJZ{
  O.~Aharony, G.~Gur-Ari, R.~Yacoby,
[arXiv:1110.4382 [hep-th]].
}

\lref\HofmanAR{
  D.~M.~Hofman, J.~Maldacena,
JHEP {\bf 0805}, 012 (2008).
[arXiv:0803.1467 [hep-th]].
}

\lref\BuchelSK{
  A.~Buchel, J.~Escobedo, R.~C.~Myers, M.~F.~Paulos, A.~Sinha and M.~Smolkin,
JHEP\ {\bf 1003}, 111  (2010).
[arXiv:0911.4257 [hep-th]].
}

\lref\DolanDV{
  F.~A.~Dolan and H.~Osborn,
[arXiv:1108.6194 [hep-th]].
}

\lref\VasilievEV{
  M.~A.~Vasiliev,
Phys.\ Lett.\ B\ {\bf 567}, 139  (2003).
[hep-th/0304049].
}

\lref\VasilievBA{
  M.~A.~Vasiliev,
In *Shifman, M.A. (ed.): The many faces of the superworld* 533-610.
[hep-th/9910096].
}

\lref\SezginRT{
  E.~Sezgin and P.~Sundell,
Nucl.\ Phys.\ B\ {\bf 644}, 303  (2002), [Erratum-ibid.\ B\ {\bf 660}, 403  (2003)].
[hep-th/0205131].
}

\lref\BashamIQ{
  C.~L.~Basham, L.~S.~Brown, S.~D.~Ellis and S.~T.~Love,
Phys.\ Rev.\ D\ {\bf 17}, 2298  (1978).
}

\lref\EastwoodSU{
  M.~G.~Eastwood,
Annals Math.\ \ {\bf 161}, 1645  (2005).
[hep-th/0206233].
}

\lref\MikhailovBP{
  A.~Mikhailov,
[hep-th/0201019].
}


\lref\solvayhigher{
Solvay workshop proceedings ,~
http://www.ulb.ac.be/sciences/ptm/pmif/HSGT.htm
}

\lref\BianchiYH{
  M.~Bianchi and V.~Didenko,
[hep-th/0502220].
}

\lref\FranciaBV{
  D.~Francia and C.~M.~Hull,
[hep-th/0501236].
}

\lref\deBuylPS{
  S.~de Buyl and A.~Kleinschmidt,
[hep-th/0410274].
}

\lref\PetkouNU{
  A.~C.~Petkou,
[hep-th/0410116].
}

\lref\BouattaKK{
  N.~Bouatta, G.~Compere and A.~Sagnotti,
[hep-th/0409068].
}

\lref\BekaertVH{
  X.~Bekaert, S.~Cnockaert, C.~Iazeolla and M.~A.~Vasiliev,
[hep-th/0503128].
}

\lref\GiombiVG{
  S.~Giombi and X.~Yin,
JHEP\ {\bf 1104}, 086  (2011).
[arXiv:1004.3736 [hep-th]].
}

\lref\GiombiYA{
  S.~Giombi and X.~Yin,
[arXiv:1105.4011 [hep-th]].
}

\lref\Evans{
N.T.Evans,
J.Math.Phys. 8 (1967) 170-185.}

\lref\KonsteinBI{
  S.~E.~Konstein, M.~A.~Vasiliev and V.~N.~Zaikin,
JHEP\ {\bf 0012}, 018  (2000).
[hep-th/0010239].
}

\lref\KochCY{
  R.~d.~M.~Koch, A.~Jevicki, K.~Jin and J.~P.~Rodrigues,
Phys.\ Rev.\ D\ {\bf 83}, 025006  (2011).
[arXiv:1008.0633 [hep-th]].
}

\lref\AnselmiBB{
  D.~Anselmi,
Class.\ Quant.\ Grav.\  {\bf 17}, 1383 (2000).
[hep-th/9906167].
}

\lref\RosensteinPT{
  B.~Rosenstein, B.~J.~Warr and S.~H.~Park,
Phys.\ Rev.\ Lett.\  {\bf 62}, 1433 (1989).
}

\lref\SezginPT{
  E.~Sezgin and P.~Sundell,
JHEP {\bf 0507}, 044 (2005).
[hep-th/0305040].
}


\lref\Vasiliev{
M.~A.~Vasiliev,
In *Shifman, M.A. (ed.): The many faces of the superworld* 533-610.
[hep-th/9910096].
}

\lref\Vasilievb{
 M.~A.~Vasiliev,
Phys.\ Lett.\ B {\bf 567}, 139 (2003).
[hep-th/0304049].
}

\lref\Vasilievc{
M.~A.~Vasiliev,
[arXiv:1203.5554 [hep-th]].
}

\lref\HolographySun{
  E.~Sezgin and P.~Sundell,
Nucl.\ Phys.\ B {\bf 644}, 303 (2002), [Erratum-ibid.\ B {\bf 660}, 403 (2003)].
[hep-th/0205131]
}

\lref\GiombiWH{
  S.~Giombi and X.~Yin,
JHEP {\bf 1009}, 115 (2010).
[arXiv:0912.3462 [hep-th]].
}


\lref\GaberdielPZ{
  M.~R.~Gaberdiel and R.~Gopakumar,
Phys.\ Rev.\ D {\bf 83}, 066007 (2011).
[arXiv:1011.2986 [hep-th]].
}

\lref\GaberdielWB{
  M.~R.~Gaberdiel and T.~Hartman,
JHEP {\bf 1105}, 031 (2011).
[arXiv:1101.2910 [hep-th]].
}

\lref\GaberdielZW{
  M.~R.~Gaberdiel, R.~Gopakumar, T.~Hartman and S.~Raju,
JHEP {\bf 1108}, 077 (2011).
[arXiv:1106.1897 [hep-th]].
}

\lref\ChangMZ{
  C.~-M.~Chang and X.~Yin,
[arXiv:1106.2580 [hep-th]].
}

\lref\ChangVK{
  C.~-M.~Chang and X.~Yin,
[arXiv:1112.5459 [hep-th]].
}

\lref\PapadodimasPF{
  K.~Papadodimas and S.~Raju,
Nucl.\ Phys.\ B {\bf 856}, 607 (2012).
[arXiv:1108.3077 [hep-th]].
}


\lref\AnninosUI{
  D.~Anninos, T.~Hartman and A.~Strominger,
[arXiv:1108.5735 [hep-th]].
}

\lref\Ng{
  G.~S.~Ng and A.~Strominger,
[arXiv:1204.1057 [hep-th]].
}

\lref\KazakovKM{
  D.~I.~Kazakov,
Phys.\ Lett.\ B {\bf 133}, 406 (1983).
}

\lref\GiombiVG{
  S.~Giombi and X.~Yin,
JHEP {\bf 1104}, 086 (2011).
[arXiv:1004.3736 [hep-th]].
}

\lref\GiombiYA{
  S.~Giombi and X.~Yin,
[arXiv:1105.4011 [hep-th]].
}



\hfill
{
PUPT-2410
}

\Title{
\vbox{\baselineskip12pt
}}
{\vbox{\centerline{  Constraining conformal field theories}
\vskip .5cm
\centerline{  with a slightly broken  higher spin symmetry }}}
\bigskip
\centerline{   Juan Maldacena$^a$ and Alexander Zhiboedov$^b$}
\bigskip

\centerline{ \it  $^a$School of Natural Sciences, Institute for
Advanced Study} \centerline{\it Princeton, NJ, USA}

\centerline{\it $^b$Department of Physics, Princeton University}
\centerline{\it Princeton, NJ, USA}

\vskip .3in \noindent

We consider three dimensional conformal field theories that have a higher spin symmetry
that is slightly broken. The theories  have a large $N$ limit, in the sense that
the operators separate into single trace and multitrace and obey the usual large $N$
factorization properties. We assume that the  spectrum of single trace operators
 is similar to the one that one gets in the Vasiliev theories. Namely, the only single
trace operators are the higher spin currents plus an additional   scalar.  The anomalous
dimensions of the higher spin currents are of order $1/N$.
Using the slightly broken higher spin symmetry we constrain the three point functions of
the theories to leading order in $N$.
We show that there are two families of solutions. One family can be realized as a theory
of $N$ fermions with an $O(N)$ Chern-Simons gauge field, the other as a $N$ bosons plus  the Chern-Simons gauge field. The family of solutions is parametrized by the
't Hooft coupling. At special parity preserving points we get the critical $O(N)$ models, both
the Wilson-Fisher one and the Gross-Neveu one.  Our analysis also fixes the on shell
three point functions of  Vasiliev's theory on $AdS_4$ or $dS_4$.


 \Date{ }



\listtoc \writetoc
\vskip .5in \noindent

\newsec{Introduction}

In this paper we study a special class of three dimensional
conformal field theories that
have a weakly broken higher spin symmetry. The theories have a structure
similar to what we expect for the CFT dual to a weakly coupled
four dimensional higher
spin gravity theory in $AdS_4$
 \refs{\Vasiliev, \Vasilievb, \Vasilievc, \SundborgWP,\wittentalk,\HolographySun,\KlebanovJA,\SezginPT,\GiombiWH}.
We   compute the leading order three point functions of the   higher spin
operators. We   use  current algebra methods.
 Our only assumption is that the correlation functions defined on the boundary of
 $AdS_4$ obey all the properties that  a boundary CFT would obey.  But we will
not need any details regarding this theory other than some general features  which
follow from natural expectations for a weakly coupled  bulk dual.
This seems a reasonable assumption for Vasiliev's theory, since Vasiliev's theory appears
to be local on distances much larger than the $AdS$ radius. This would imply that the usual
definition of boundary correlators is possible \refs{\GiombiWH,\GiombiVG,\GiombiYA}.
 In order to apply our analysis to
Vasiliev's theory we need to make the assumption that these boundary correlators can be defined
and that they obey the general properties of a CFT. Thus we are assuming AdS/CFT, but we are
not specifying the precise definition of the boundary CFT.
  Since our assumptions are very general,
they also apply to theories involving $N$ scalar or fermions fields coupled to
$O(N)$ or $SU(N)$ Chern-Simons gauge fields \refs{\GiombiKC,\AharonyJZ}. So our methods are also useful for
computing three point functions in these theories as well.

Our assumptions are the following\foot{ These assumptions are not all independent
 from each other, but we will not give the minimal set.}.
 We have a CFT with a unique stress tensor and has
a large parameter $\tilde N$. In the Chern-Simons examples $\tilde N \sim N$. In the
  Vasiliev gravity theories, $1/\tN \sim \hbar $ sets  the bulk coupling constant of the theory.
  We then assume that the spectrum of operators has the structure
of an approximate Fock space, with single particle states and multiparticle states.
The dimensions of
the  multiparticle states   are given by the sum of the dimensions
of their  single particle constituents   up to small $1/\tilde N$ corrections.
 This Fock space
should not be confused with the Fock space of a free theory in three dimensions. We should think of this Fock space as the Fock space of  the weakly coupled four dimensional gravity theory.
 In order to avoid this confusion we will call the ``single particle states'' ``single trace'' and the
 multiparticle states ``multiple trace''. In the Chern-Simons gauge theories this is indeed the case.
 We also assume that the  theory
has the following spectrum of single trace states.
 It has a single spin two conserved
current. In addition, it has a sequence of approximately conserved currents $J_s$, with
$s =4,6,8, \cdots $. These currents are approximately conserved, so that their twist differs from one by
a small amount of order $1/\tN$
\eqn\smallviol{
\tau_s = \Delta_s  - s =   1 + O({1 \over \tN}) .
}

In addition,  we have one single trace scalar
operator.
All connected correlators of the single trace operators scale as $\tilde N$. This
includes the two point function of the stress tensor.
 We also  assume that the spectrum
of single trace operators is such that the higher spin symmetry can be broken only by double
trace operators
 via effects of order $1/\tilde N$. In particular, we assume that there are no twist three single
 trace operators in the theory.


With these assumptions,
we will find that the three point functions in these theories, to leading order in
$\tilde N$,  are constrained to lie on a one parameter family
\eqn\corrf{
\langle J_{s_1} J_{s_2} J_{s_3} \rangle = \tilde N \left[
{\tla^2 \over 1 + \tla^2 }  \langle J_{s_1} J_{s_2} J_{s_3} \rangle_\bos +
{ 1 \over 1 + \tla^2 }\langle J_{s_1} J_{s_2} J_{s_3} \rangle_\fer +
{ \tla \over 1 + \tla^2 }\langle J_{s_1} J_{s_2} J_{s_3} \rangle_\odd \right]
}
where the subindices $~_\bos$ and $~_\fer$ indicate the results in the theory of
a single real boson or a single Majorana fermion.
 The subindex $_\odd$ denotes
an odd structure  which will be defined more clearly below. Here $\tla$ is a parameter
labeling the family of solutions of the current algebra constraints. More precise statements will
be made below, including the precise normalization of the currents.

The class of theories for which our assumptions apply includes Vasiliev higher spin
theories in $AdS_4$ with higher spin symmetry broken by the boundary conditions
\Vasiliev .
It also applies for theories containing $N$ fermions \GiombiKC\  or $N$ bosons  \AharonyJZ\  interacting with
an $SO(N)$ or $U(N)$ Chern-Simons gauge field.
We call these theories quasi-fermion  and quasi-boson   theories respectively.
In such theories $1/\tilde N \propto 1/N$ is the small parameter. In addition,
$\tilde \lambda  \sim  \lambda = N/k $ is an effective 't Hooft coupling in these theories.
 We emphasize that the analysis here is only based on the symmetries and it covers
 both types of theories, independently of any conjectured dualities between them.
 Of course,  the results we obtain are consistent with the proposed dualities between these
 Chern-Simons theories and Vasiliev's theories \refs{\KlebanovJA,\HolographySun,\GiombiWH,\GiombiKC,\AharonyJZ}.

We can take the limit of large $\tla $ in \corrf\ and we find that the the correlators
of the quasi-fermion theory go over to those of the critical $O(N)$ theory. And
we have a similar statement for the quasi-boson theory.

 Our analysis is centered on studying the   spin four single trace operator  $J_4$. We write
 the most general form for its divergence, or lack of conservation. With our assumptions this
 takes the schematic form
 \eqn\basicA{
  \partial . J_4 = a_2 J J' + a_3 JJ'J''
  }
  where on the right hand side we have products of   two or three single trace
  operators, together with derivatives sprinkled on the right hand side.
  The coefficients $a_2$ and $a_3$ are small quantities of order $1/\tN$ and $1/\tN^2$ respectively.
  We will be able to use this  approximate conservation law in the expression for the
  three point function in order to get \corrf . Note that in the case that $J_4$ is exactly conserved,
  we simply have free boson or free fermion correlators \MaldacenaJN .

We should emphasize that our discussion applies only to the special theories in
\refs{\GiombiKC,\AharonyJZ}, but not to more general large $N$ Chern-Simons matter theories.
The special
feature that we are using is the lack of single trace operators of twist three. Such
operators
 can appear in the divergence of the spin four current \basicA . This can  give rise to an anomalous dimension for the  higher spin currents
already at the level of the classical theory (or large $\tilde N $ approximation).
In the theories in \refs{\GiombiKC,\AharonyJZ}, we
do not have a single trace operator   that
can appear in the right hand side of \basicA .
In the language of the bulk theory, we have the pure higher spin theory without extra matter\foot{
By ``matter'' we mean extra multiplets under the higher spin symmetry.
The scalar field of the Vasiliev theory is part of the pure higher spin theory. }.
In particular, the bulk theory lacks the matter fields that could give a mass to the higher
spin gauge fields via the Higgs mechanism already at the classical level.
Thus the higgsing is occurring via quantum effects involving
two (or three) particle states  \GirardelloPP .

Our analysis can also be viewed as an on shell analysis of the Vasiliev theory with
$AdS_4$ asymptotic boundary conditions. If the higher spin symmetry is unbroken, then
we can use \MaldacenaJN\ to compute all correlators, just from the symmetry. In this
paper we also use an on shell analysis, but for  the case that the higher spin symmetry
is broken. As it has often been emphasized, on shell results   in  gauge theories
can be simpler than what the fully covariant formalisms would suggest.


For a more general motivational introduction,  see appendix G.

  The paper is organized as follows.

In section two we discuss the most general form of the divergence of the spin four current.

In section three we present several facts about three point functions which are necessary for the later analysis.

In section four we explain how one can use the slightly broken higher spin symmetry to fix the three point functions.

In section five we present the results and explain their relation to known microscopic theories.

In section six we present conclusions and discussions.

Several appendices contain technical details used in the main body of the paper.


\newsec{Possible divergence of the spin four operator  }

\subsec{Spectrum of the theory }

We consider theories with a large $\tN$ expansion. We do not assume that
$\tN $ is an integer. We assume that the set of operators develops the structure
of a Fock space for large $\tN$. Namely, we can talk about single particle
operators and multiparticle operators. In the case of the Chern-Simons matter theories
discussed in \refs{\GiombiKC,\AharonyJZ}, these correspond to single sum and multiple sum operators (sometimes
called single trace or multitrace operators).
The spectrum of single trace operators includes a conserved spin two current, the
stress tensor, $J_{2 \, \mu \nu}$. We will often suppress the spacetime indices
and denote operators with spin simply by  $J_s$. We have
approximately conserved single trace operators $J_s$, with $s=4,6,8, \cdots $. These operators
have twists $\tau = 1 + O(1/\tN)$.
In addition we have a single scalar operator. We will see that the dimension
of this operator has to be either one or two. We denote the first possibility as
$j_0$ and the second as $\tilde j_0$. The theory that contains $\tilde j_0$ is called the
quasi-fermion theory. The theory with $j_0$ is called the quasi-boson theory.
The theory might contain also single particle
operators with  odd spins. For simplicity, let us  assume that these are not
present, but we will later allow their presence and explain that the correlators of even spin currents
are unchanged.

An example of a theory that obeys these properties is a theory with $N$ massless
Majorana fermions
interacting with an $O(N)$ Chern-Simons gauge field at level $k$ \GiombiKC . In this theory
the scalar is $\tilde j_0 = \sum_i \psi_i \psi_i $, which has dimension two, at leading order in $N$ for any
$\lambda = N/k$.
The name quasi-fermion was inspired by this theory, since we start from fermions and
the Chern-Simons interactions turns them into non-abelian anyons, which for large $k$,
are very close to ordinary fermions. Our discussion is valid for any theory whose single particle spectrum
was described above. We are just calling ``quasi-fermion'' the case where the spectrum includes a scalar
with dimension two, $\tilde j_0$.

A second example is a theory with $N$ massless real
 scalars, again interacting with an $O(N)$ Chern-Simons gauge field at level $k$ \AharonyJZ .
 This theory also allows the presence of a $ ( \vec \phi .\vec \phi)^3$ potential while preserving
 conformal symmetry, at least to leading order in $N$.  As higher orders in the $1/N$ expansion are taken into account
 this coefficient is fixed, if we want to preserve the conformal symmetry \AharonyJZ .  Here we will only do computations
 to leading order in $N$, thus, we have two parameters $N/k$ and the coupling of
 the $ ( \vec \phi .\vec \phi)^3$ potential. We call this case the quasi-boson theory. Again we will not use any of the
 microscopic details of its definition. For us the property that defines it is that the scalar has dimension one,
 $j_0$.

 A third example is the critical $O(N)$ theory (as well as interacting UV Gross-Neveu fixed point). Namely, we can have $N$ free scalars,
 perturbed by a potential of the form $  ( \vec \phi .\vec \phi)^2$, which flows in the IR to a new
 conformal field theory (after adjusting the coefficient of the mass term to criticality).
 This is just the usual large $N$ limit of the Wilson-Fisher fixed point. This theory
 has no free parameters. Here the scalar operator $\tilde j_0 \sim \phi_i \phi_i $ has dimension two. It starts with dimension one in the UV but it has dimension two in the
 IR. The UV theory has a higher spin symmetry. In the IR CFT this symmetry is broken
 by $1/N$ effects. This theory is in the family of what we are calling the quasi-fermion case.

 A fourth example is a Vasiliev theory in $AdS_4$ (or $dS_4$) with general boundary conditions
 which would generically break the higher spin symmetry. Here the bulk coupling is
 $\hbar \sim 1/\tN $. Depending
 on whether the scalar has dimensions one or two, we would have a quasi-fermion or quasi-boson case.

 We should emphasize that the theories we call quasi-fermion or quasi-boson case are not specific microscopic
 theories.  They are {\it any} theory
 that obeys our assumptions, where the scalar has  dimension two or one respectively.

  \subsec{Divergence of the spin four current}

  Let us consider the spin four current $J_4$. We consider the divergence of this
  current. If it is zero, then we have a conserved higher spin current and
   all correlators of the currents are as in a theory of either free bosons or
  free fermions  \MaldacenaJN . Here we consider the case that this divergence is nonzero.
  Our assumptions are that the current is conserved in the large $\tN$ limit.
  This means that in this limit $J_4$ belongs to a smaller multiplet than at finite $\tN$. At finite $\tN$,  $J_4$ combines with another operator to form a full massive multiplet. More precisely,
  it combines with the operator that appears in the right hand side of $\nabla . J_4 = \partial_\mu J^\mu_{\ \ \nu_1 \nu_2 \nu_3} $.
  In other words, $\nabla . J_4$, should be a conformal primary operator in the large $N$ limit
\refs{\HeidenreichXI,\GirardelloPP},  see appendix A.
  $\nabla . J_4 $  should be a twist three, spin three primary operator.
  According to our assumptions there are no single particle operators of this kind.
   Note that in general matter Chern-Simons
  theories,
  such as theories with adjoint fields, we can certainly have
   single trace operators with twist
  three and spin three. So, in this respect,  the theories we are considering are very
  special.
  In our case, we can only have two particle or three particle states with these quantum numbers.

Let us choose the metric
\eqn\metric{
ds^2 = dx^+ dx^- + dy^2
}
and denote the indices of a vector by $v_\pm $, $v_y$.

  Let us see what is the most general expression we can write down for
   $(\nabla . J) _{---}$.
  Since the total twist is three, and the total spin is three we can only make this
  operator out of the stress tensor, $J_2$,  and the scalar field. Any attempt to include a higher spin
  field would have to raise the twist by more than three.
  A scalar field can only appear if
  its twist is one or two. Note that we cannot have two stress tensors. The reason
  is that we cannot make a twist three, spin three primary out of two stress tensors.
  Naively, we could imagine an expression like $J_{2 \, y-} J_{2 \, --}$. But this cannot
  be promoted into a covariant structure, even if we use the $\epsilon$ tensor.\foot{We
  could have promoted it if we had two different spin two currents: $\eps_{- \mu \nu} J^{\mu}_{\ -} \tilde J^{\nu}_{\ -}$ }

  Let us consider first the case that the scalar has twist two,   $\tilde j_0$.
   The most general operator that  we can write down is
  \eqn\sayco{
  \partial_\mu J^\mu _{\ ---} =  a_2 \left( \pa_{-}\tilde j_0 j_2 - {2 \over 5} \tilde j_0 \pa_{-} j_2 \right).
   }
 Here we are denoting by $j_s = J_{s \, -\cdots -}$, the all minus components of $J_s$.
 In \sayco\ we have used the fact that the right hand side should be a conformal primary in
 order to fix the relative coefficient. If we started out from the free fermion
 theory, then this structure would break parity, since $\tilde j_0$ is parity odd. Then we have
  $a_2 \propto N/k$, at least for large $k$. In general $\tilde j_0$ does not have well defined parity and
  the theory breaks parity.
 If  we had  the critical O(N) theory,  then \sayco\ is perfectly consistent
 with parity, since in that case $\tilde j_0$ is parity even.

 Let us now consider the case where we have a scalar of twist one, $j_0$.
 Now  there are more conformal primaries that we can write down
\eqn\divscalarNice{\eqalign{
\pa_{\mu} j^{\mu}_{\ \ ---} &=  a_2 \eps_{- \mu \nu}  \left[8 \pa^{\mu} j_{0} \pa^{\nu} j_{--} - 6   \pa^{\mu} j_{0} \pa_{-} j^{\nu}_{ \ -} - 5 \pa_{-} \pa^{\mu} j_0 j^{\nu}_{\ -} - j_0 \pa_{-} \pa^{\mu} j^{\nu}_{\ -} \right]\cr
&+ a_3 \left[ j_0 j_0 \pa^3 j_0 - 9 j_0 \pa j_0 \pa^2 j_0 + 12 \pa j_0 \pa j_0 \pa j_0\right] \cr
&+a'_3  \left[ \pa j_{2} j_0 j_0 - 5 j_{2} \pa j_0 j_0 \right].
}}
We have only written combinations which are conformal primaries. We have also denoted
$\partial \equiv \partial_-$.
The analysis of the broken charge conservation identities will relate $a_3$ and $a'_3$ leaving us
with only two parameters (besides $\tN$). This agrees with the two parameters in the large $N$ limit in
the boson plus Chern-Simons theories of \AharonyJZ .

Here we have concentrated on the case of $J_4$. Let us briefly discuss the situation
for higher spin currents.
 We focus,  as usual,  on the  all minus component of the current $\pa_\mu J^{\mu}_{\ --...-}$.
 This operator has twist $\tau = \Delta -S = 3 $ and spin $s - 1$.
 Let us examine possible double particle operators that can appear.
 The minimum twist of a  double trace operator is $2 = 1 + 1$. We should make up
  the twist by considering other components, or derivatives other than $\partial_-$, which has twist zero. All of these should arise from a rotationally invariant structure involving the flat space metric or the $\epsilon$ tensor.
    The only structure that can raise the  twist by one   is $\epsilon_{- \mu \nu}$. For the quasi-fermion theory we can also use the scalar operator of twist $2$, $ \tilde j_0$,  and one of the twist one currents.

Matching the scaling dimensions in  $\pa_\mu J^{\mu}_{s} \propto J_{s_1} J_{s_2}$
(with derivatives sprinkled on the right hand side)
with all minus indices  leads to
\eqn\matching{\eqalign{
s + 2 = (s_1 + 1)+ (s_2+1) + n_{\pa}
}}
where $n_{\pa} \geq 0$ is the number of derivatives which raise the dimension. Thus, we get an inequality
\eqn\matchingb{
s \geq s_1 + s_2, ~~~~~~ s > s_{1} , ~s_2  ~,~~~ ~~ {\rm double ~ trace}
}
where we show that  that $s > s_1,s_2 $ as follows.
 For $s_1 =s$, the only operator with the right twist
  would  be $j_0 J_{s\, y-\cdots-}$, but this is not
 really a spin $s-1$ operator, namely it does not come from any covariant structure.
 \matchingb\  is a constraint on the spins of the operators that can appear in the
 divergence of a current of spin $s$.

At the level of triple trace operators the product of three operators has already twist three. So the only structure which is allowed is $\pa_-$ by the twist counting. Matching the dimensions in  $\pa_\mu J^{\mu}_{s} \propto J_{s_1} J_{s_2} J_{s_3}$  we get
\eqn\matchingC{\eqalign{
s + 2 = (s_1 + 1) + (s_2+1) + (s_3 + 1) + n_{\pa}
}}
and the constraint
\eqn\matchingd{
s \geq s_1 + s_2 + s_3 + 1, ~~~~ {\rm triple ~ trace}
}
Twist counting prohibits having the product of more than three operators.

Now let us comment on the scaling of the coefficients in \basicA\  with $\tilde N$.
Let us normalize the scaling of single particle operators so that their  connected
$n$ point
functions scale like $\tilde N$.
Then if we consider a three point correlator of a given current with the two currents
that appear in the right hand side of its divergence we get
\eqn\scale{
\tN \sim \pa_{\mu} \la J_s^{\mu}(x)  J_{s_1}(x_1)  J_{s_2}(x_2)  \ra = a_2 \la J_{s_1}(x) J_{s_1}(x_1)  \ra \la J_{s_2 }(x)J_{s_2}(x_2)  \ra \sim a_2 \tN^2
}
with derivatives sprinkled on the right hand side.
Thus, we get that $a_2 \propto {1 \over \tN}$. For $a_3$ the same argument leads to $a_3 \propto {1 \over \tN^2}$. This scaling is the only
one that is consistent with the $\tN$ counting and is such that it leads to non-zero terms in the leading contribution.

\newsec{Structures for the three point functions }

In this section we  constrain the
structure of three point functions. When we have exactly
conserved currents the possible three point functions
were found in   \GiombiRZ\ (see also \CostaMG \MaldacenaJN ).
 They were found by imposing
 conformal symmetry and current conservation.
 The three point functions
were given by three possible structures. One structure arises in the free fermion
theory and  another arises  in the free boson theory. We call these the fermion and boson
structures respectively. Finally there is a third ``odd'' structure which does
 not arise in a free theory.  For twist one fields this structure is   parity odd
 (it involves an epsilon tensor).
 However, for correlators of the form  $ \langle \tilde j_0 J_{s_1} J_{s_2}\rangle$,
 the fermion structure is parity odd (it involves an epsilon tensor)
 and the ``odd'' structure is parity even.
  This is due to the fact that  $\tilde j_0$ is parity odd in the free fermion
theory.  Alternatively, if we
 assign parity minus to $\tilde j_0$ and parity plus to all the
twist one operators, then the ``odd'' structures
always violate parity\foot{This is not always the natural parity assignment. For example, in the critical
$O(N)$ theory $\tilde j_0$ has parity plus and the theory preserves parity. In this theory,
we have only the ``odd'' structure for the
correlators of the form $ \langle \tilde j_0 J_{s_1} J_{s_2}\rangle$. }. The reader should
think that when we denote a structure as ``odd'' we simply mean ``strange'' in the sense that
it does not arise in a theory of a free boson or free fermion.

In our case, the currents are not conserved, so we need to revisit these
constraints.  For example, the divergence of a current can produce a double trace operator.
If these contract with the two remaining operators, as in \scale , we get a term
that is of the same order in the $1/\tN$ expansion as the original three point function.
 Notice that   only double trace operators can contribute in this manner to the
 current non-conservation of a three point function. We emphasize that we are
 computing these three point functions to leading order in the $1/\tN$ expansion, where
 we can set their twist to be one. All statements we make in this section are about the
 structure of correlators to leading order in the $1/\tN$ expansion.


We will show below that even correlation functions (fermion and boson)
stay the same and all new structures appear in the odd piece.
Consider the three point function of twist one operators $\la J_{s_1} J_{s_2} J_{s_3} \ra$, with $s_i \geq 2$. Let us say that $s_1$  is larger or equal than the other two spins.
  Then  the $J_{s_2}$ and $J_{s_3}$
currents are conserved inside this three point function, since
  the spins appearing in the
 divergence of a current are always strictly  less than those of the current itself
 \matchingb  . On the other hand, in order to get a non-zero contribution we would
 need to contract $J_{s_1}$ with one of the two currents that appears in the
 right hand side of the divergence of $J_{s_2}$ or $J_{s_3}$.
Thus, we can impose current conservation on $J_{s_2}$ and $J_{s_3}$  for this three
point function.
Let us consider the parity even structures first.
As we discussed in \MaldacenaJN ,  for two operators of the same twist and one
 conserved current, say $j_{s_3}$,
 we have the most general even structure
\eqn\genstr{
\langle O_{s_1} O_{s_2} j_{s_3} \rangle \sim  { 1 \over |x_{12} |^{ 2 \tau_0 -1}  |x_{23}| |x_{13}|}   \sum_{l=0}^{ \,  {\rm min} [ s_1 , s_2 ]}   P_{3}^{2 l }  \left[
\langle j_{s_1-l} j_{s_2 -l } j_{s_3} \rangle_\bos + \langle j_{s_1-l} j_{s_2 -l} j_{s_3} \rangle_\fer  \right]
}
where $P_i  $ are as is \GiombiRZ .  We get this result by considering  the
light cone limit between $\underline{O_{s_1} O_{s_2}}$ and imposing the conservation of $j_{s_3}$ \MaldacenaJN .
We then take light cone limit $\underline{j_{s_1}j_{s_2}}$ and impose
conservation of $j_{s_3}$. Then we take the light cone limit $\underline{j_{s_1}j_{s_3}}$ and impose conservation of $j_{s_2}$. From these two operations we would conclude that
\eqn\gensoltwo{\eqalign{
\langle j_{s_1} j_{s_2} j_{s_3} \rangle &=  { 1 \over |x_{12} |  |x_{23}| |x_{13}|}   \sum_{l=0}^{   \,  {\rm min} [ s_1 , s_2 ]}   P_{3}^{2 l }  \left[ b_l \langle j_{s_1-l} j_{s_2 -l } j_{s_3} \rangle_\bos + f_l \langle j_{s_1-l} j_{s_2 -l} j_{s_3} \rangle_\fer \right], \cr
\langle j_{s_1} j_{s_2} j_{s_3} \rangle &=  { 1 \over |x_{12} |  |x_{23}| |x_{13}|}   \sum_{l=0}^{   \,  {\rm min} [ s_1 , s_3 ]}   P_{2}^{2 l }  \left[ \tilde b_l \langle j_{s_1-l} j_{s_2} j_{s_3 - l} \rangle_\bos + \tilde f_l \langle j_{s_1-l} j_{s_2 } j_{s_3 - l} \rangle_\fer  \right].
}}
The only consistent solution is $b_0 = \tilde b_0$, $f_0 = \tilde f_0$ and $b_l$ and $f_l$ with $l \neq 0$ are equal to zero. This can be seen by taking the light cone limit
in $x_{12}$ first, which sets to zero all terms of the form $P_3^{ 2 l}$, with $l>0$, as
well as the fermion terms. In the second line only the boson structures survive, but only the $l=0$ structure is the same as the one surviving in the first line. This shows that
all $\tilde b_l =0$ for $l> 0$. Repeating this argument we can show it for the other
cases.

We can now consider also the case when one of the particles has spin zero, or is $j_0$.
Then any of the expressions in \gensoltwo\ only allows the $l=0$ term. Thus, the even
structures with only one $j_0$ are the same as in the free boson theory (the free
fermion ones are zero).


For the odd structure the situation is more tricky.
Inside the triangle,  $ s_i \leq s_{i+1} + s_{i-1}$  for $i=1,2,3$,
we have the structures that we had before, since \matchingb\ does not allow any
of the three currents to have a non-zero divergence.
However, outside the triangle we  have
new structures that obey \scale\ with a non-zero double trace term.
Precisely the existence of these new structures make the whole setup consistent.
The current non-conservation identity has the form of the current conservation one
but with a non-zero term in the right hand side. Since outside the triangle we had
no solutions of the homogeneous equations, this guarantees that the solutions are
uniquely fixed in terms of the operators that appear in the right hand side of
the conservation laws. Thus, we have unique solutions for these structures.

One interesting example of this phenomenon is the correlator
\eqn\specialWIb{
\la J_4 J_2 j_0 \ra_{\odd , \, nc} \propto  a_2 { S_3 Q_1^2 \over |x_{12}| |x_{13}| |x_{23}|} \left[ Q_1 Q_2 + 4 P_3^2  \right].
}
where $P_i$ and $Q_i$ are defined in \GiombiRZ .
This odd correlator would be zero if all currents were conserved. However, using the
lack of conservation of the $J_4$ current, \divscalarNice , we can derive \specialWIb .
Clearly only $a_2 $ contributes to it.







As an another example, consider $\langle J_4 J_2 \tilde j_0 \rangle$.
Here we can have structures that are parity odd and parity even, the fermion and the
``odd'' structure respectively.
Due to the form of the current non-conservation of $J_4$, \sayco , we get
\eqn\specialWI{
\la J_4 J_2 \tilde{j}_0 \ra_{\odd , \,  {\rm nc}} \propto a_2
 { Q_1^2 \over |x_{12}| |x_{13}| |x_{23}|} \left[ P_3^4 - 10 P_3^2 Q_1 Q_2 - Q_1^2 Q_2^2  \right]
}
This ``odd''  structure would vanish if $J_4$ were exactly conserved.
Of course, we also have the structure that we get in the free fermion theory for
these case, which is parity odd (while \specialWI\ is parity even).
 In fact, any correlator of two twist one currents and one
$\tilde j_0$, which is parity odd will have the same structure as in the free fermion
theory since \sayco\ (or its higher spin versions), will not modify it. On the
other hand, the parity even ones can be modified. Again we find that a structure that
was forced to be zero when the current is exactly conserved can become non-zero
when the current is not conserved.

Finally, we should mention that any correlator that involves a current and two
scalars is uniquely determined by conformal symmetry. In this case the current is
 automatically conserved.
Of course, the three point function of three scalars is also unique.

In summary, we constrained  the possible structures for various three point functions.
These are the boson, fermion and odd structures. When the operator $\tilde j_0$ is involved
we can only have fermion or odd structures.
In the next section we will constrain the relative coefficients of all of these three point functions.

\newsec{Charge non-conservation identities}

\subsec{General story}

We use the following technique to constrain the three point function.
 We  start from a three point function $\la O_1 O_2 O_3 \ra $.
  We then insert a $J_4$
current and take its divergence, which gives us an identity of the form
\eqn\diem{  \eqalign{
\langle \nabla . J_4(x)  O_1 O_2 O_3 \rangle = &
a_2 \langle   J J'  O_1 O_2 O_3 \rangle  + a_3 \langle J J' J''  O_1 O_2 O_3 \rangle \cr
~ \sim  &  a_2   \langle J O_1 \rangle \langle J' O_2 O_3 \rangle
+ a_3 \langle J  O_1 \rangle \langle J'  O_2 \rangle \langle J''  O_3\rangle
+ {\rm permutations}
}}
This equation is schematic since we dropped derivatives that should be sprinkled on
the right hand side. The two point functions in the right hand side  are
 non-zero only if $J$ or $J'$ is the
same as one of the operators $O_i$.   Thus the  the right hand side  is  non-zero only when  any
of the operators $O_i$ is the same as one of the currents that appears in the right
hand side of the divergence of $J_4$ \basicA \sayco \divscalarNice .
 We have only considered disconnected contributions
in the right hand side because those are the only ones that survive to leading order
in $1/\tN$. Here we used the scaling of the coefficients $a_2$ and $a_3 $ with $\tN$
given in \scale .

Given this equation, we can now integrate  over $x$ on the left and right hand sides.
We integrate over a region which includes the whole space  except for little spheres, $S_i$,
 around each of
the points $x_i$ where the operators $O_i$ are inserted.
The left hand side of \diem\ contributes only with a boundary term of the form
\eqn\equb{
\sum_{i=1}^3 \langle \int_{S_i} n^j J_{j ---}  O_1  O_2 O_3  \rangle
}
where the integral is over the surface of the little spheres or radii $r_i$
around each of the points and $n^j$ is the normal vector to the spheres.

\ifig\Spheres{The action of the charge is given by the integral of the conserved current over a little sphere around the operator insertion. To derive the pseudo-conservation identity we are integrating the divergence of the current over all  space except for these spheres. The
boundary terms give the pseudo-charges, which have the same expression in terms of
the current as in the conserved case.   } {\epsfxsize2.2in\epsfbox{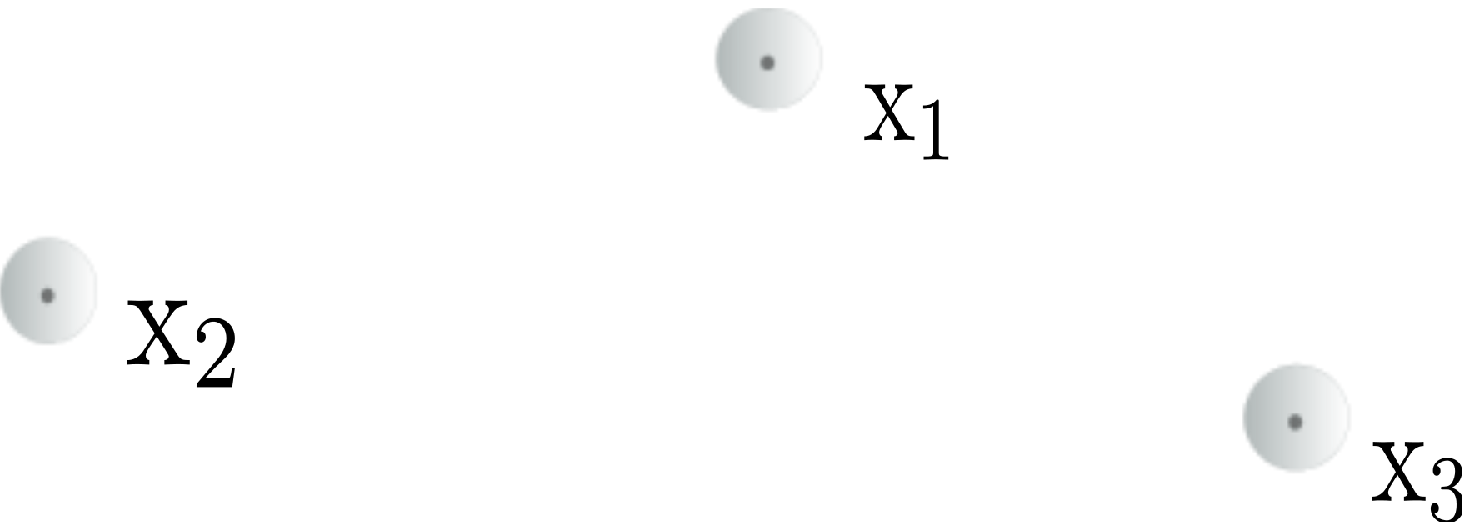}}

If the current $J_4$ were exactly conserved, these integrals would give the charge acting
on each of the operators. In our case, the charges are not conserved and the integrals
may depend on the radius of the little spheres. This dependence can  give rise to
divergent terms going like inverse powers of the radius of the spheres. These terms diverge
when $r_i\to 0$. These divergent terms should precisely match similar divergent terms
that arise in the integral of the right hand side of \diem . After matching all the divergent terms
we are left with the finite terms in the $r_i \to 0 $ limit.  These also have to match between
the left and right hand sides. Demanding that they match we will get interesting constraints.
Notice that, at the order we are working, we do not get any logarithms of $r_i$ since
the anomalous dimensions of operators start at higher order of the $1/\tilde N$ expansion.
Thus, the separation of the finite and the divergent terms is always unambiguous.

Thus, we define a pseudo-charge  $Q$ that acts on the operators by selecting the
finite part of the above integrals in the small radius limit
\eqn\actfi{
 [Q, O(0)] =   \left.  \int_{|x|=r} n^j J_{j ---}  O (0) \right|_{{\rm finite~as~} r\to 0 }
 }

 This action of this pseudo-charge on single trace operators is determined by the three point
 functions we discussed above. It is also constrained by twist and spin conservation to
 have a similar structure to the one we had for absolutely conserved currents. For example,
 on  the twist one operators  we have
  \eqn\expcst{
 [ Q , j_s] =  \sum_{s' = 0}^{s+3}   c_{s,s'}   \partial^{ s - s' + 3}  j_{s'}
 }

 In concrete computations we found it useful to work in the metric \metric\ and to cut
 out little ``slabs''  of width $\Delta  x^+ $ around the operators, instead of cutting out
 little spheres.

\ifig\Slabs{ Instead of spheres as in \Spheres , we can cut out little
 slabs of width $\Delta x^+$  around the insertion point of every operator. The charge is given by integrating current over $x^{-}$ and $y$ at the edges of these slabs. This
 simplifies some computations compared to \Spheres . } {\epsfxsize2.2in\epsfbox{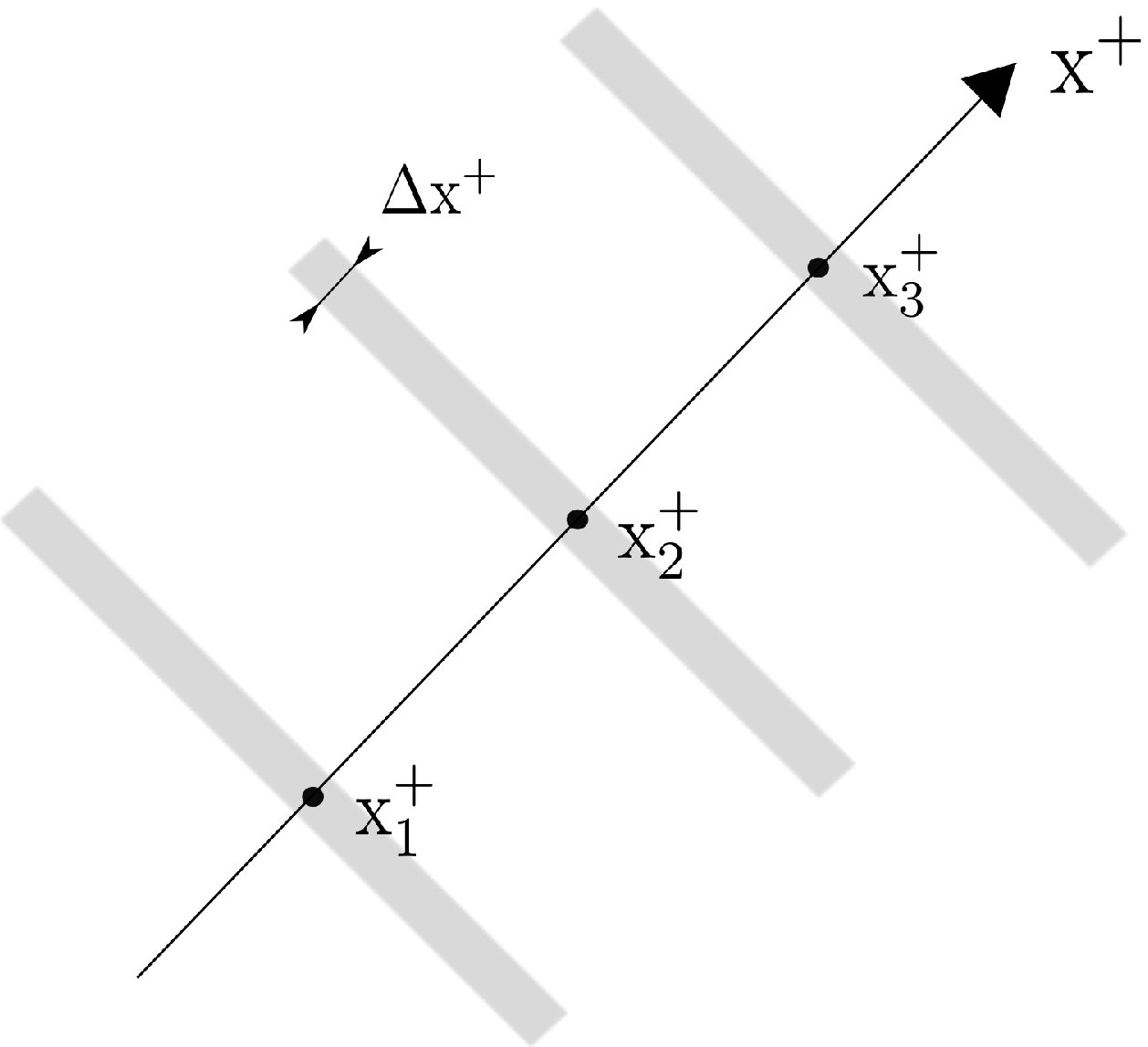}}

  The advantage is that then the integral involves the all minus component
 of the current. In addition,  minus and $y$-derivatives can be integrated by parts or
  be pulled out of the integral
%

\eqn\chargesnon{
Q_s (x^{+}) = \int d x^{-} d y \ j_{s} (x^-,x^+,y).
}

The action of the pseudo-charge on $O_i$ comes from
 three point functions of the form
$\langle J_4 O_i O_k \rangle$. As we mentioned the structures that  arise in the boson
or fermion theories will continue to produce three point functions where the charges are
conserved.
 The odd correlators can give us something new. The odd structures
involving all twist ones fields, such as $\langle j_4 j_{s_2} j_{s_3} \rangle$,
are parity odd and do not contribute to the action of $Q$.
 This can be seen by setting
  $y_2 = y_3 = 0$. Then the fact that the three point function is odd under the
   parity $y_1 \to - y_1$ implies that the integral over $y$ in \chargesnon\
    must vanish. This implies
that commutator must vanish also for arbitrary $y_2 \neq y_3$ since the two point function structures that could possibly contribute  do not vanish for $y_2 = y_3$.
When one of the operators $O_i$ has twist two,  we can get non-vanishing contributions
to the action of the pseudo-charge   from odd structures.

In conclusion, after integrating \diem\ we get an expression of the form
\eqn\intexp{\eqalign{
\langle [Q,O_1] O_2 O_3 \rangle + {\rm cyclic} = &
 \int d^3 x
 \left[ a_2   \langle J(x)  O_1 \rangle \langle J'(x) O_2 O_3 \rangle + \right.
\cr
~ + & \left. a_3 \langle J(x)  O_1 \rangle \langle J'(x)  O_2 \rangle \langle J''(x)  O_3\rangle  + {\rm permutations } \right]_{\rm finite}
}}
where the operators $O_i$ are evaluated at $x_i$. The integral is over the full $R^3$ after
subtracting all the divergent terms that can arise around each of the points $x_i$.
This is the main identity that we will use to relate the various three point functions
to each other. We can call it a pseudo-conservation of the pseudo-charges.

\subsec{Constraints on three point functions with non-zero even spins  }

In this section we will consider the constraints that arise on three point
functions of operators with spins $s_i\geq 2$. We will consider all their
indices to be minus. So we take the operators $O_i$ in \diem\ to be $j_{s_i}$, with
$s_i \geq 2$.

In this case the action of $Q$ can only produce other twist one fields, which are only single particle
states.  As we mentioned above the action of $Q$ is determined by
three point functions of the form
$\langle j_4 j_{s_1} j_{s_2} \rangle$ or $\langle j_4 j_{s_1} j_0 \rangle$.
 Only the even structures contribute to the charges, and these
are the same (up to overall coefficients)
as in the free theories, thus the action of the pseudo-charge is well defined and produces
\eqn\normal{
[Q , j_s ] = c_{s,s-2} \partial^5 j_{s-2} + c_{s,s} \partial^3  j_s + c_{s,s+2 } \partial  j_{s+2}
}
where $\partial = \partial_-$.
We can always choose the following normalization conditions
\eqn\normco{
c_{2,0} =1~,~~~~~~~~~ c_{2,4} =1 , ~~~ c_{4,4} =1, ~~~~ c_{4,6} =1 , ~~~ c_{s,s+2 } =1 , s \geq 2
}
This can be done in all cases, pure fermions, pure bosons, or the interesting theory we
are considering. We are not setting the normalization of two point function to one.
 The two point functions are not
used for the time being. The stress tensor is normalized in the canonical way.
Notice that, since  the normalization of stress tensor is fixed, $c_{2,4}$ does not depend on the normalization of $j_4$ and
is fixed by conformal invariance to 3 (recall $[Q_{s}, j_2] = (s-1) j_s$ \MaldacenaJN ).
Thus, we fix it to one by the rescaling freedom in the definition of $Q_4 \to { 1 \over 3 }  Q_4$.
Then we fix $c_{4,4} =1$ by rescaling $j_4$ itself. And $c_{s,s+2 } =1$ by rescaling  $j_{s+2}$.

Now, in our case, it is clear that if all $s_i \geq 4$, then the right hand side of \intexp\ vanishes.
In addition, if one or more, of the $s_i$ is equal to two, then the following happens. In that case the two
current terms in \sayco, \divscalarNice\ could  contribute, since we can contract the $J_2$ in the divergence of $J_4$ with
the $s_i$ that is equal to two.
It is useful to recall the two point functions of various components of the stress tensor
\eqn\twoppp{\eqalign{
\la J_{--}(x) J_{--} (0)\ra &\propto {(x^{+})^4 \over (x^{+} x^{-} + y^2)^{5}} = {1 \over 4!} \pa_{-}^4 {1 \over x^+ x^- + y^2},  \cr
\la J_{--}(x) J_{-y} (0)\ra &\propto 2 {(x^{+})^3 y \over (x^{+} x^{-} + y^2)^{5}} = {1 \over 4!} \pa_{-}^3 \pa_{y} {1 \over x^+ x^- + y^2}, \cr
\la J_{-y}(x) J_{-y} (0)\ra &\propto  - {(x^{+})^2 (x^+ x^- - 3 y^2 ) \over (x^{+} x^{-} + y^2)^{5}} = {1 \over 4!} \left(\pa_{-}^2 \pa_{y}^2 {1 \over x^+ x^- + y^2} - 2 \pa_{-}^2 {1 \over (x^+ x^- + y^2)^2} \right).
}}
where the equations are true up to a  normalization factor common to  all three equations.

If we look at the first term in the right hand side of \intexp\ in the case that $O_1 = j_2$, then
we can use the above two point function \twoppp . We can integrate by parts all derivatives
so that they act on the two point function. Then we can write them as derivatives acting on
$x_i$ and pull them out of the integral\foot{If we work with the ``slabs'' described in
figure \Slabs, together with the usual $i \epsilon $ prescription,
 it is clear that these operations do not produce boundary terms since we only
have $\partial_-$ or $\partial_y$ derivatives.}.

Now the  result depends on whether we are dealing with the quasi-fermion or quasi-boson cases.
In the quasi-boson case, it is possible to check that the particular combination of currents that appear in
the two current term in \divscalarNice\  is, after integrating by parts,
  $\partial_y J_{2 \, --} - \partial_- J_{2 \, y-} $. When this is contracted with $j_2 = J_{2\, --}$ we get zero after
  using \twoppp . Thus, the right hand side of \intexp\ vanishes in the quasi-boson case.

  In the quasi-fermion case, we end up having to compute (the $\partial_-^5$ derivative)
  of an integral of the form
\eqn\fermextra{
\int d^3 x {1 \over |x - x_1|^2} \la \tilde{j}_0 (x) O_2(x_2) O_3(x_3) \ra \propto  \langle j_0^\eff(x_1)  O_2(x_2) O_3(x_3) \rangle
}
The
 factor ${1 \over |x - x_1|^2} $  is exactly the one that makes the integral conformal covariant. It gives a result
 that effectively  transforms  as the three point function with  the insertion of an operator of dimension
 $\Delta =1$ at $x_1$. We have denoted this in terms of an effective operator, $j^{\eff}_0$, of dimension one.
   We do not have any real operator of this kind in the theory, this is just a mnemonic to help us keep track
   of the resulting integral.
 The net result can be expressed by saying that the action of $Q$ on $j_2$ in the quasi-fermion theory
 can also produce a $j_0^\eff $. In the free boson theory the action of $Q$ on $j_2$ produces $j_0$ (plus other
 things). This is not true in the free fermion theory, which does not contain a $j_0$. However, in the
 quasi-fermion theory, with the pseudo-conservation property \sayco \intexp , we get terms in these identities
 that would be identically to what we would get if we had allowed a $j_0^\eff $ in the action of $Q$ on $j_2$.
  This term is crucial in order to allow a free boson structure in the pseudo-conservation identities.

In other words, we define a new effective pseudo-charge $\hat Q$, which has the same
action as $Q$ on all currents with spin $s>2$, but it is
\eqn\acto{
[\hat Q , j_2 ] =  c_{2,0} \partial^5  j_0^{eff}  + c_{2,2} \partial^3  j_2 + c_{2,4} \pa j_4
}

The net result is that when we consider the pseudo-conservation
 identities acting on $\langle j_{s_1}
j_{s_2} j_{s_3} \rangle$ becomes identical to a charge conservation identity for $\hat Q$.
Note also that after adding $j_0^{eff}$ in \acto\ the action of $\hat Q$ is essentially the same
as the action of $Q$ in the quasi-boson theory. Thus, we can treat these two cases in
parallel, after we remember that $j_0^{eff}$ is not a real operator but just the integral in
\fermextra .

Next we write all twist one three point functions   as
\eqn\writen{
\la j_{s_1} j_{s_2} j_{s_3} \ra = \alpha_{s_1,s_2,s_3} \la j_{s_1} j_{s_2} j_{s_3} \ra_{\bos}+ \beta_{s_1,s_2,s_3} \la j_{s_1} j_{s_2} j_{s_3} \ra_{\fer} + \gamma_{s_1,s_2,s_3} \la j_{s_1} j_{s_2} j_{s_3} \ra_{\odd}
}

Here the boson and fermion ones are the three point functions
 for a single real boson and a single Majorana  fermion, in the normalization
of the currents set by \normco .
The normalization of the odd piece is fixed so that the identities we describe below are true.
In the quasi-fermion theory we are also including in \writen\ the case with
 $ \langle j_0^{eff} j_{s_1} j_{s_2} \rangle$, where only the boson and odd structures
 are non-zero.


We now consider the charge conservation identities for various cases.
An important property of these identities is that each identity separates into three
independent equations which relate only the boson structures to each other, the fermion
structures to each other and the odd structures to each other.
Each  equation involves  sums of objects of the form $ \partial_i^n \la j_{s_1} j_{s_2} j_{s_3} \rangle_{{\rm str}} $, where ${\rm str}$ runs over
boson, fermion and odd.    These equations are such  that the coefficient in
front of each of the terms is fixed relative to all the other coefficients. In other words, only an overall
constant is left undetermined. See appendix B. This was proven for the boson and
fermion terms in \MaldacenaJN . We have not proved this for the odd case.
 However, given
the existence of the Chern-Simons matter theories \refs{\GiombiKC,\AharonyJZ} we know
that at least one solution definitely exists! For
 low values of the spins we have explicitly analyzed the equations and the solution is definitely unique, in the sense that all relative coefficients are fixed.
We think that this is likely to be true for all cases.

Focusing first on the equations that constrain the boson structures  we get that
the equations arising from pseudo-charge conservation are
\eqn\wardeq{ \eqalign{
Q \la j_2j_2j_2\ra  :&  ~~~ \tilde{c}_{2,2} \alpha_{222} = \alpha_{224} = \alpha_{022}
\cr
Q \la  j_2 j_2 j_4\ra  :&  ~~~ \tilde{c}_{2,2}   \alpha_{224} = \tilde{c}_{4,4}  \alpha_{224} = \alpha_{244} =  \tilde{c}_{4,2} \alpha_{222} =   \alpha_{226 } = \alpha_{024}
\cr
Q \la  j_2 j_4 j_4\ra  :&  ~~~ \tilde{c}_{2,2}  \alpha_{244} = \tilde{c}_{4,4}  \alpha_{244} = \alpha_{444} =  \tilde{c}_{4,2} \alpha_{224} =   \alpha_{246 } = \alpha_{044}
\cr
Q \la  j_2 j_2 j_6\ra  :&  ~~~ \tilde{c}_{2,2}  \alpha_{226} =   \tilde{c}_{6,6} \alpha_{226} = \alpha_{246} =  \tilde{c}_{6,4} \alpha_{224} =
   \alpha_{228} = \alpha_{026}
}}
and we can continue in this way.
We are defining $\tilde c_{s,s'}$ to be the ratio
\eqn\sacn{
\tilde c_{s,s'} = { c_{s,s'} \over c_{s,s'}^{\rm free~ boson} }
}
with both in the normalization \normco .

We now can start solving \wardeq .
 We see that the equations \wardeq\ fix all the $\alpha$'s in terms of $\alpha_{222} $
and the $c$'s. In addition, we get multiple equations for the same $\alpha$'s.
This fixes the $\tilde c$'s.
For example, we start getting things like
\eqn\listofq{\eqalign{
 \alpha_{224} & =  \tilde{c}_{2,2} \alpha_{222}
 \cr
\alpha_{244} & =   \tilde{c}_{4,2} \alpha_{222} ~,~~~~  \tilde{c}_{2,2}  =  \tilde{c}_{4,2} ~,~~~~~ \alpha_{226} =  \tilde{c}_{4,2} \alpha_{222}
\cr
\alpha_{444} & =  \tilde{c}_{4,2}  \tilde{c}_{2,2} \alpha_{222} ~,~~~~   \tilde{c}_{2,2}  =  \tilde{c}_{4,4} ~,~~~ \alpha_{244} =  \tilde{c}_{4,2} \alpha_{222} ~,~~~\alpha_{246} =  \tilde{c}_{4,2}  \tilde{c}_{2,2} \alpha_{222} .
}}
where each line in \listofq\ comes from the corresponding line in \wardeq .
Using the fact that $ \tilde{c}_{4,4} = 1$
we see that $ \tilde{c}_{2,2} =1$, and also $\tilde{c}_{4,2} =1$, and so on. So all the
$\tilde c_{s,s'} =1$.
We also find that all the $\alpha$'s are also fixed to be equal to $\alpha_{222}$.

If we did the same for the free fermions we would also obtain the same pattern if we define
  \eqn\sacn{
\tilde c_{s,s'} = { c_{s,s'} \over c_{s,s'}^{\rm free ~ fermion} }
}
Then we get that all $\beta_{s_1,s_2,s_3} = \beta_{222}$ and all $\tilde c_{s,s'} =1$.
 One subtlety is that we have
defined the $\tilde c's$  differently
 for the bosons than for the fermions. So we can only hope to get both structures
 present only if
\eqn\workcond{
c_{s,s'}^{\rm free~ boson} = c_{s,s'}^{\rm free ~fermion}
}
This can be checked by using form factors (see Appendix C). Of course, the mere existence
of the Chern-Simons matter theories implies that this is true.

 The conclusion from this analysis is  that all the $\alpha$'s are equal to  $\alpha_{222}$.
 Analogously, all $\beta$'s are equal to  $\beta_{222}$.

 We will also need the coefficients of the two point functions $n_s$.
 In a theory of a single
 free boson or single free fermion, with the normalization conditions \normco , these are
 given by $n_s^{bos}$ and $n_s^{fer}$ respectively. We do not need their explicit forms,
 but they can be computed in the free theories.
 We can now determine the two point functions in the full theory
 by demanding that the stress tensor Ward identities are obeyed.
 In other words, from the previous discussion we know that
  $\alpha_{2ss} = \alpha_{222}$ and  $\beta_{2ss} = \beta_{222}$.
 In addition, the Ward identity of the stress tensor relates this to the two point function.
  More precisely,
 \eqn\normtwo{
  n_s = \alpha_{222} n_s^{bos} + \beta_{222} n_s^{fer}
  }
  Notice that $n^{bos}_2 = n^{fer}_2$ according to formulas (5.1) and (5.7) of
  \MaldacenaNZ  \foot{ There the formula for the Dirac fermion is written. Here   we consider a Majorana one.} . Thus we find
%
\eqn\normtwoB{
\tilde{n}_2 \equiv  { n_2 \over n_2^{bos} } =  \alpha_{222} + \beta_{222}
}

Notice that the analysis so far is equivalently valid for the theories of quasi-bosons and the theory of quasi-fermions. Of course, in the latter case, whenever a $j_0$ appeared, it should be interpreted as
 $j_0^\eff$.

Thus, so far, we have written all correlators in terms of three undetermined coefficients $\alpha_{222}$,
$\beta_{222}$ and $\gamma_{222}$. As an aside we should note that when we solve the equations for
the $\gamma_{s_1s_2 s_3}$ we need to use some of the odd correlators that are outside the triangle. These
are possible thanks to the non-conservation of the currents.

\subsec{Closing the chain. The Quasi-fermion case.  }

In this section we consider current conservation identities in the case that
 one of the operators is a   twist two fields such as $\tilde j_0$ or
$J_{2\, -y}$.

Let us start by discussing the possible action of the pseudo-charge
$Q$ on $J_{-y}$. All single trace operators that could appear are already fixed by
the correspondent three point functions to be the same as in free fermion theory.
The most general expression involving double trace terms
takes the form
\eqn\consi{
[ Q, J_{- y } ] = \partial^4 \tilde j_0 + \pa^3  J_{- y}  + \pa J_{4 ---y} + {\chi \over \tilde N}  j_2 j_2 .
}
The double trace term comes with an ${1 \over \tilde N}$ factor because it enters in the pseudo-conservation
identity with the $\la j j\ra \la j j\ra \sim \tilde N^2$ factor.

 We have normalized $\tilde j_0$ by setting the first coefficient to one\foot{
 This is not possible for  the critical $O(N)$ theory since in that case parity prevents
 $\tilde j_0$ in the right hand side. We will obtain this case as a limit. It can also be analyzed
 directly by considering the action of $Q$ on $\tilde j_0$, etc. }.
 One can check that $\chi $ should be set to zero by considering $\la j_2 j_2 J_{-y}\ra$ pseudo-conservation identity  \foot{ This double trace structure
 can arise in a covariant way as $[Q , J_{ 2 \, \mu \nu} ] = \eps_{- \mu \rho} J^{\rho}_{\ -} J_{\nu -} + \eps_{- \nu \rho} J^{\rho}_{\ -} J_{\mu -}$.}.

In addition,  we write the possible correlators of $\tilde j_0$ in terms of free fermion correlators,
introducing a $\beta_{\tilde 0 s_1 s_2 }$. This is the coefficient that multiplies
$\langle \tilde j_0 j_{s_1} j_{s_2}\rangle _\fer $, which is
the correlator  in a theory of a single Majorana fermion in the
normalization of $\tilde j_0$ in \consi .
Note, that these free fermion correlators are parity odd.
We also introduce a $\gamma_{\tilde 0 s_1 s_2 }$ which multiply  the ``odd'' structures,
which are  parity even structures involving
$\tilde j_0$. These structures are more subtle since they can be affected by the violation of current conservation,
as in \specialWI .

An additional issue we should discuss is the type of contribution we expect from the right hand side of
\intexp\  when the operator is $J_{- y}$. The same reasoning we used around \fermextra , together with \twoppp\ tells
us that we also effectively produce a $ j_0^\eff $. Thus, the net result is that
 we simply should add terms involving $j_0^{eff}$ in
\consi , and treat the charge as conserved. This is necessary for getting all the identities to work.
We then find that $\beta_{\tilde 0 s_1 s_2} = \beta_{222}$  and $\gamma_{\tilde 0 s_1 s_2 } =
\gamma_{222}$.

Interestingly, in this case,
 in order to  satisfy the pseudo-charge conservation identities for the odd part, we need both
$\tilde j_0$ as well as $j_0^{eff}$.


Let us now consider the three point function $\langle j_0^{eff} j_2 j_2 \rangle$. We see
that its definition via \fermextra\ involves a three point function given by
$\beta_{\tilde 0 22}$ and $\gamma_{\tilde 0 22 }$. But we have just
fixed these coefficients.
 Thus, going back to the first line \wardeq , we note that
$\alpha_{022} $ is really the coefficient of $\langle j_0^{eff} j_2 j_2 \rangle_\bos $. We get
this structure from doing the integral \fermextra\  with the ``odd'' structure for the three
point function.
Using that  $\alpha_{0 22} $ was fixed to $\alpha_{222}$ we get the equation
   \eqn\odiot{
    \alpha_{222}  =x_2 a_2 n_2 \gamma_{\tilde 0 22} =\tilde x_2 a_2 ( \alpha_{222} + \beta_{222} ) \gamma_{222}
    }
    Similarly, the odd structure gives
    \eqn\idenob{
    \gamma_{222} =x_1 a_2 n_2 \beta_{\tilde 0 22} =\tilde x_1 a_2 ( \alpha_{222} + \beta_{222} ) \beta_{222}
    }
      where $x_1$ and $x_2$ are two
       calculable numerical coefficients, given by doing the integral in
    \fermextra , etc. We used \normtwoB\ and $\tilde x_i = x_i/n_2^\bos $.

    These are two equations for four unknowns ($\alpha_{222}, ~\beta_{222},~\gamma_{222}, ~a_2)$.
    So we have two undetermined coefficients which are $\tilde{N}$ and $\tilde{\lambda}$. We can define     $\tilde N = \tilde n_2 = \alpha_{222} + \beta_{222}  $ and $ \tilde \lambda =  a_2 \tilde N / \sqrt{\tilde x_1 \tilde x_2}  $.
    With these
    definitions we  get
    \eqn\fins{
    \alpha_{s_1 s_2 s_3} =\tilde{N} { \tilde{\lambda}^2 \over (1 + \tilde{\lambda}^2 )} , ~~~\beta_{s_1 s_2 s_3} = \tilde{N} { 1 \over        (1 +  \tilde{\lambda}^2 ) } ~,~~~~
    \gamma_{s_1 s_2 s_3} = \tilde{N} { \tilde{\lambda} \over (1 +  \tilde{\lambda}^2 ) }.
    }
     Here we are considering three twist one fields.    Similar equations
    are true for correlators involving one $\tilde j_0$ field
    \eqn\fintwi{ \beta_{s_1 s_2\tilde 0} = \tilde{N} { 1 \over        (1 +  \tilde{\lambda}^2 ) } ~,~~~~
    \gamma_{s_1 s_2 \tilde 0 } = \tilde{N} { \tilde{\lambda} \over (1 +  \tilde{\lambda}^2 ) } .
    }
  We will discuss the correlation functions that involve more than one $\tilde j_0$ separately.
In both equations, \fins \fintwi, we can choose the ($\tilde \lambda$ independent) numerical normalization
of the odd correlators so that we can replace $\sim $ by an equality.

 The final conclusion is that we expressed all the correlators of the currents in terms of just two parameters
 $\tilde N$ and $\tilde \lambda$.

Since the fermions plus Chern-Simons  theory has precisely two parameters, $N$ and $k$ we conclude
 that we exhausted all the constraints. Our analysis was based only on general symmetry consideration and
 does not allow us to find the precise relation between the parameters. However, in the 't Hooft limit we expect\foot{For the theory considered in \GiombiKC\ $f_0 = 2$ and $d_1 = {\pi \over 2}$.}
 \eqn\expo{
 \tilde N = N f(\lambda) ~,~~~~~~~~~~~~ \tilde \lambda = h(\lambda) = d_1 \lambda + d_3 \lambda^3 + \cdots
 }
 where $f(\lambda) = f_0 + f_2 \lambda^2 + \cdots $.   Where we used the symmetry of the theory under
 $\lambda \to - \lambda$ (or $k \to - k$), together with parity. Note that the function $f$ encodes how the
 two point function of the stress tensor depends on $\lambda$.


\subsec{Closing the chain. The Quasi-boson case.  }

In the quasi-boson theory  we will again consider the charge non-conservation identity on  $\la j_2 j_2 J_{y-} \ra$.
Again we need to write the most general action of $Q$ on $J_{3 -}$ involving double trace terms. It takes the form
\eqn\consiB{\eqalign{
[ Q, J_{- y } ] &= \partial^4 \pa_{y} j_0 + \pa^3 J_{- y } + \pa J_{---y} \cr
&+{ \chi_{1} \over \tilde N} j_2 j_2 + { \chi_{2} \over \tilde N}  \pa^2 j_0 j_2 + { \chi_{3} \over \tilde N} \pa^4 j_0 j_0
}}
We have normalized $j_0$ by setting the first coefficient to one\foot{Again, this is not possible
in the critical $O(N)$ Gross-Neveu theory. But we will obtain this case as a limit. It can also
be analyzed directly through a slightly longer route. }.

Notice also that $\chi_i \sim O({1 \over \tilde N})$ to contribute at leading order in $\tilde N$. We fix $\chi_1 = 0$
as in the case of fermions. The presence or absence of $\chi_{2,3}$ is not important for any of the arguments.

In addition, we need to consider the contribution of $J_{-y}$ to the right hand side of \intexp .
In other words, we will need to consider the right hand side of \intexp\ when the operator
$Q_1$ is $J_{-y}$.
Using \twoppp , we find (derivatives of) an integral of the form
\eqn\bosextra{
\int d^3 x {1 \over |x - x_1|^4} \la j_0 (x) O_2(x_2) O_3(x_3)  \ra \sim \la  \tilde j_0^\eff(x_1) O_2(x_2) O_3(x_3)  \rangle
}
Here we used that, again, this is a conformal integral which behaves as a correlator with a scalar of weight
$\Delta =2$ at $x_1$. We have denoted this by introducing a ficticious operator $\tilde j_0^\eff $. This is
not an operator that exists in the theory, but it is appearing in three point functions in the same
way as an operator of this form.
Thus, the net effect of the action of $Q$ on $J_{-y}$ includes also this operator $\tilde j_0^{eff}$ in the
right hand side of \consi .

With all these features, we now have a situation which is rather similar to the one we had in the quasi-fermion
case and we can relate the different coefficients
\eqn\idenobB{\eqalign{
\beta_{222} = & y_1 a_2 n_2 \gamma_{0 22} = \tilde y_1 a_2  ( \alpha_{222} + \beta_{222} ) \gamma_{222}.
\cr
\gamma_{222}  = & y_2 a_2  n_2 \alpha_{0 22} =\tilde y_2  a_2  ( \alpha_{222} + \beta_{222} ) \alpha_{222}.
}}
Again $y_1$ and $y_2$ are some numbers. Defining again
$\tilde N = \tilde n_2 = (\alpha_{222} +
\beta_{222}) $ and $\tilde \lambda = a_2 \tN / \sqrt{\tilde y_1 \tilde y_2}  $ we get
\eqn\finsB{
\alpha_{s_1 s_2 s_3} =\tilde{N} {1 \over (1 +  \tilde{\lambda}^2 )} , ~~~\beta_{s_1 s_2 s_3} = \tilde{N} { \tilde{\lambda}^2 \over (1 +  \tilde{\lambda}^2 ) } ~,~~~~
\gamma_{s_1 s_2 s_3} = \tilde{N} { \tilde{\lambda} \over (1 + \tilde{\lambda}^2 ) }.
}
 Here at least two of the spins should be bigger than zero,  $s_i \geq 2$.  We will discuss the correlation functions that involve more than one $ j_0$ separately.

\subsec{Three point functions involving scalars. The quasi-fermion case. }

Fixing the three point functions in the scalar sector involves several new subtleties that were absent before.
We describe them in detail in the Appendix D and here sketch the method and present the results.

We have already discussed
how   to fix three point functions which include one scalar operator. We used
the pseudo-conservation identity on
 $\langle J_{-y}  j_{s_1} j_{s_2} \rangle $ to get \fintwi .

To proceed it is necessary to specify how $Q$ acts on $\tilde j_0$. As explained in Appendix D the result is
\eqn\actonscalar{
[Q, \tilde j_0] =\pa^3 \tilde j_0  + {1 \over 1 + \tilde \lambda^2} \pa \left[ \pa_{-} j_{- y} - \pa_{3} j_{--} \right]
}
here the interesting new ingredient is a ${1 \over 1 + \tilde \lambda^2}$ factor. This is
obtained by inserting arbitrary coefficients and fixing them by analyzing the $\la j_4 j_2 \tilde j_0 \ra$
three point function.


To fix the correlators with two scalars we consider the pseudo-charge conservation identity
that is generated by acting with $Q$ on
$\la \tilde j_ 0 \tilde j_ 0 j_s \ra$. This identity
involves tricky relations (see Appendix D) between different three point functions.
After the dust settles we get that
\eqn\twoscalars{\eqalign{
\beta_{s \tilde 0 \tilde 0} &= \beta_{2 2 2}  ~,~~~~~~~~~~~~
\gamma_{s \tilde 0 \tilde 0} = \gamma_{2 2 2}
}}
The  last step is to consider the action of $Q$ on
 $\la \tilde j_ 0 \tilde  j_0 \tilde  j_0 \ra$ WI to get
\eqn\lastferm{
 \gamma_{\tilde 0 \tilde 0 \tilde 0} = 0.
}
Note that $\beta_{\tilde 0 \tilde 0 \tilde 0} = 0$ by definition, since this correlator vanishes in
the free fermion theory (due to parity).

From these three point function it is also possible to extract the normalization of the
two point function $ \la \tilde j_0 \tilde j_0 \rangle $. This is two point function is
related by a Ward identity to
$\la j_2 \tilde j_0 \tilde j_0 \ra$. We then get
\eqn\normtwo{
{\tilde n_{\tilde 0 } } \equiv { n_{\tilde 0 } \over n_{\tilde 0 }^{\rm free~fermion}  } = \beta_{2\tilde 0 \tilde 0 } = \beta_{222} = \tilde N { 1\over 1 + \tilde \lambda^2 }
}

This can be used, together with \sayco ,  to compute the anomalous dimension for the
spin four current, as explained in appendix A.

\subsec{Three point functions involving scalars. The quasi-boson case. }

For the quasi-boson sector the story is almost identical. We put details of the analysis in the Appendix E and here
again sketch the general idea and present the results.

We have already discussed above how
to fix three point functions which include one scalar operator.
Then the charge conservation identity on
$\langle j_2 j_{s_1} j_{s_2} \rangle $ fixes
\eqn\onescqb{\eqalign{
\alpha_{0 s_1 s_2} &= \alpha_{2 2 2} ~,~~~~~~~~
\gamma_{0 s_1 s_2} = \gamma_{2 2 2}.
}}
For the action of the charge on the scalar we get
\eqn\actonscalarbos{
[Q, j_0] =\pa^3 j_0  + {1 \over 1 + \tilde  \lambda^2} \pa j_2.
}
By considering the pseudo-conservation of   $\la j_s  j_0 j_0 \ra$ when $s > 2$ we get
\eqn\twoscalarsB{\eqalign{
\alpha_{s 0 0} &= \beta_{2 2 2}~,~~~~~~~~~
\gamma_{s  0  0} = \gamma_{2 2 2}.
}}
We now need to consider the pseudo-conservation identities for   $\la j_2 j_0 j_0\ra$ and
 $\la j_0 j_0 j_0\ra$.
A new feature of these two cases is that
 the triple trace terms in \divscalarNice\  contribute.
 Analyzing these we obtain that \twoscalarsB\ is also true for $s=2$.
The triple trace terms contribute as follows.
Let us first consider the pseudo-conservation identity for  $\la j_0 j_0 j_0 \ra$.
The triple trace non-conservation term takes the form
\eqn\anom{\eqalign{
a_3 n_0^3 \sum_{i=1}^{3} \pa^3_{i}  \la j_0 (x_1) j_0 (x_2) j_0 (x_3) \ra.
}}
where $n_0$ is the coefficient in the two point function for the scalar.
Thus, we get the equation
\eqn\neweq{
\alpha_{000} ={1 \over 1 + \tilde \lambda^2} \alpha_{222} +  z_1 a_3 n_0^3
}
where $z_1$ is a numerical constant and $b_3$ is the coefficient in \divscalarNice .
Notice that the double trace deformation does not influence this computation. This fact is established in the Appendix C.

The triple trace term in the pseudo-conservation identity for  $\la j_2 j_0 j_0 \ra$
is
\eqn\anomtwo{\eqalign{
a'_3  n_2 n_0^2 \pa^5_{1}  \la j_0 (x_1) j_0 (x_2) j_0 (x_3) \ra.
}}
leading to
\eqn\neweqB{
\alpha_{000}  ={1 \over 1 +  \tilde \lambda^2} \alpha_{222} +  z_2 a'_3 n_0^2 n_2 + ...
}
with $z_2$ a numerical constant
 and the  dots stands for the contribution of the double trace non-conservation piece whose
 coefficients are already known.

Importantly, we conclude that two triple trace deformations are not independent. In other words,
$a_3$ and $a'_3$ are related by equating \neweqB\ and \neweq .
 Thus, we recover the known counting of marginal deformation of free boson in $d=3$, namely there are two parameters. Microscopically, one corresponds to the Chern-Simons coupling
 and the second one to adding a $ ( \vec \phi. \vec \phi)^3$ operator.

On the Vasiliev theory side,  the freedom to add  this $\phi^6$ deformation translates
into the fact that we can choose a non-linear boundary conditions for the scalar
which preserves the conformal symmetry. These were discussed in a similar situation in
\HertogDR . In our context, we have a scalar of mass $ ( m R_{AdS})^2 =-2)$ which
at infinity decays as $\phi = \alpha/r + \beta/r^2$. Then the boundary
condition that corresponds to adding the $\lambda_6 \phi^6$ deformation is
$\beta = \lambda_6 \alpha^2 $ \HertogDR .

Also the whole effect of the presence of the
 triple trace deformations, at the level of three point functions,
  is to change $\la j_0 j_0 j_0 \ra$.


Finally, the two point function of $j_0$ can be fixed by using the usual stress tensor
Ward identity via $\la j_2 j_0 j_0 \ra $. We obtain
\eqn\twopfo{
 { \tilde n_0 } \equiv { n_0 \over n_0^{\rm free~boson} } = \alpha_{222} =\tN { 1 \over 1 + \tilde \lambda^2 }
 }
Recall that the normalization of $j_0$ was given by setting $c_{2,0}=1$ in \normco .

\subsec{Comments about higher point correlation functions}

We can wonder whether we can determine higher point correlation functions.
 It seems possible to
use the same logic. Namely, inserting $ \nabla . J_4$ into an $n$ point function and  then integrating  as in
\intexp . This relates the action of $Q$ on an $n$ point function to integrals of disconnected
correlators. These integrals involve also $n$ point functions. (Recall that for three point
functions the integrals involved other three point functions).
 When the charge is conserved this is expected to fix the connected
correlation uniquely to that of the free
theory. This was done explicitly in \MaldacenaJN\
 for the action of $Q$ on $\langle j_0 j_0 j_0 j_0 \rangle $.
Now that the right hand side is non-zero, we still expect this to fix uniquely the correlator, though  we have
not tried to carry this out explicitly. It is not totally obvious that this will fix the correlators because
the integral terms also involve $n$ point functions.
But it seems reasonable to conjecture that this procedure would fix
the leading order connected correlator for all $n$ point functions of single particle operators. It would be
interesting to see whether this is indeed true!

\newsec{Final results }

In this section we summarize the results for the three point functions.

The normalization of the stress tensor is the canonical one. The normalization of the
charge is $Q = { 1 \over 3} \int j_4 $. This sets $c_{2,4}=1$ in \normal  .
 The normalization of $j_4$ is fixed by setting $c_{4,4}=1$. Then all other operators
 are normalized by setting $c_{s,s+2}=1$. The operator $\tilde j_0$ is normalized by
setting $c_{2,\tilde 0} =1$ in \consi .

We have the two point functions
\eqn\twopnot{
\la j_s (x_1) j_{s} (x_2) \ra = n_{s} {(x_{12}^{+})^{2 s} \over | x_{12} |^{4 s + 2}} .
}
For three point functions
\eqn\threepnot{\eqalign{
\la j_{s_1} (x_1) j_{s_2} & (x_2) j_{s_3} (x_3) \ra =
\alpha_{s_1 s_2 s_3} \la j_{s_1} (x_1) j_{s_2} (x_2) j_{s_3} (x_3) \ra_\bos +
\cr
& +  \beta_{s_1 s_2 s_3} \la j_{s_1} (x_1) j_{s_2} (x_2) j_{s_3} (x_3) \ra_\fer
 +
\gamma_{s_1 s_2 s_3} \la j_{s_1} (x_1) j_{s_2} (x_2) j_{s_3} (x_3) \ra_\odd
}}
where the $_\bos$ and $_\fer$ denote the three point functions in the theory of free boson and free fermion in the normalization of currents described above. Their functional form can be found in \GiombiRZ . The odd generating functional for the spins inside the triangle can be found in appendix B of \MaldacenaJN .
 Outside the triangle the odd correlation functions are the ones that satisfy the double trace deformed non-conservation equations.
  We do not
   know their explicit form in general but nevertheless we know that they exist and know how the dependence on the coupling will enter. We fix the numerical normalization of the odd pieces
  to be such that the pseudo-charge conservation identities are obeyed.  In a similar fashion we define
  the correlators involving a $\tilde j_0$ ($j_0$) operator, except that there is no free boson (fermion) structure.

\subsec{Quasi-fermion theory }

The  interacting theory two point functions are given by
\eqn\twopfe{\eqalign{
n_s &=  \tilde{N} {\tilde \lambda ^2 \over 1 + \tilde \lambda ^2}  n_{s}^{\rm free~boson} + \tilde{N} { 1 \over  1 + \tilde \lambda ^2 }  n_{s}^{\rm free~fermion} = \tilde{N} n_{s}^{\rm free~boson} , ~~~~ s \geq 2, \cr
n_{\tilde 0}  &=  \tilde{N} { 1 \over  1 + \tilde \lambda ^2 }  n_{\tilde 0 }^{\rm free~fermion}
}}
where $ n_{s}^{\rm free~boson} $  and  $  n_{s}^{\rm free~fermion}$
are two point functions computed in the theory of single free boson or single fermion with normalization of operators such that \normco\ (and \consi ) holds. We also used in the first line the fact that $n_{s}^{\rm free~boson} =  n_{s}^{\rm free~fermion}$ in the normalization that we adopted. This is explained in the appendix C.

The three point functions in the interacting theory are then given by
\eqn\finsfe{\eqalign{
\alpha_{s_1 s_2 s_3} &= \tilde{N}  {\tilde \lambda ^2 \over 1 + \tilde \lambda ^2}
 , ~~~~~\beta_{s_1 s_2 s_3}  = \tilde{N}  {1  \over 1 + \tilde \lambda ^2}
  ~,~~~~~
\gamma_{s_1 s_2 s_3} = \tN  {\tilde \lambda  \over 1 + \tilde \lambda ^2}
\cr
\beta_{s_1 s_2 \tilde 0}  &= \tilde{N} { 1  \over 1 + \tilde \lambda ^2}
, ~~~~~~~~
\gamma_{s_1 s_2 \tilde 0} = \tilde{N} {\tilde \lambda  \over 1 + \tilde \lambda ^2}
\cr
\beta_{s_1 \tilde 0 \tilde 0}  &= \tilde{N}  {1 \over 1 + \tilde \lambda ^2}
 \cr
\gamma_{\tilde 0 \tilde 0 \tilde 0} &= 0
}}
All coefficients not explicitly written do not appear because
 there is no corresponding structure.
The two parameters are defined as follows. We take the stress tensor to have a canonical normalization.
We then set
\eqn\paramfe{\eqalign{
\tilde N &={  n_2 \over n_2^{\rm free~boson} }
, \cr
\tilde \lambda ^2 &= {\alpha_{222}
 \over \beta_{222}
 }.
}}
From the bounds on energy correlators discussed in \MaldacenaJN\ it follows that $\tilde \lambda^2 \geq 0$.
Note that one would then find that $\tilde \lambda  \propto a_2 \tilde N$, with $a_2$ defined in \sayco .
We have also computed the anomalous dimension of the spin four current (see appendix A)
\eqn\anmdima{
\tau_4 -1  = { 32 \over 21 \pi^2 } { 1 \over \tilde N }  {\tilde  \lambda^2 \over (1 + \tilde \lambda^2 ) }
}

\subsec{Quasi-boson theory }

The interacting theory two point functions are given by
\eqn\twopbo{\eqalign{
n_s &=  \tilde{N}  {1   \over 1 + \tilde \lambda ^2}
 n_{s}^{\rm free ~boson}
  + \tilde{N}  {\tilde \lambda^2  \over 1 + \tilde \lambda ^2}
  n_{s}^{ \rm free ~fermion} = \tilde{N} n_{s}^{\rm free ~boson} ~, ~~~~ s \geq 2, \cr
n_{ 0}  &=  \tilde{N}  {1  \over 1 + \tilde \lambda ^2}
 n_{0}^{\rm free ~  boson}
}}
and the three point  functions are
\eqn\finsbo{\eqalign{
\alpha_{s_1 s_2 s_3} &= \tilde{N}  {1 \over 1 + \tilde \lambda ^2}
 , ~~~~~\beta_{s_1 s_2 s_3}  = \tilde{N}  {\tilde \lambda^2  \over 1 + \tilde \lambda ^2}
  ~,~~~~~
\gamma_{s_1 s_2 s_3} = \tN  {\tilde \lambda  \over 1 + \tilde \lambda ^2}
\cr
\alpha_{s_1 s_2  0}  &= \tilde{N} { 1  \over 1 + \tilde \lambda ^2}
, ~~~~~~~~
\gamma_{s_1 s_2  0} = \tilde{N} {\tilde \lambda  \over 1 + \tilde \lambda ^2}
\cr
\alpha_{s_100}  &= \tilde{N}  {1 \over 1 + \tilde \lambda ^2}
 \cr
\alpha_{ 0  0  0} &=  \tilde{N} { 1 \over  (1 + \tilde \lambda ^2)^2 }  +  z_1 a_3  n_0^3
}}
were we have separated out the correlators involving the scalar.
Two of the parameters are defined by  (taking into account that stress tensor has a canonical normalization)
\eqn\parambo{\eqalign{
\tilde N &= { n_2 \over n_2^{\rm free~boson} } , \cr
\tilde \lambda^2 &= {\beta_{222}   \over \alpha_{222} }.
}}
Again one can see that $\tilde \lambda^2 \geq 0$ from the bounds on energy correlators discussed in \MaldacenaJN.
A third parameter can be introduced which  is the combination involving $a_3$ that
shifts the three scalar correlator in \finsbo .
Again we find that $\tilde \lambda  \propto a_2 \tN$ in \divscalarNice .
 Thus we find a three parameter family of solutions, as expected to leading order in the
 large $N$ limit.

\subsec{The critical point of $O(N)$ }

The critical point of $O(N)$ is a theory that we obtain from the free boson theory by adding a $j_0^2$ interaction
and flowing to the IR, while tuning the mass of the scalars to criticality. Then the operator $j_0$ in the
IR gets to have dimension two, for large $N$.
 Thus, we can view it as a $\tilde j_0$ scalar operator of dimension two.
This is a theory that fits into the quasi-fermion case. Namely the divergence of $J_4$ is given by
\sayco . By the way, if we assume that the IR limit has a scalar operator of dimension different from one, then
since \sayco\ is the only expression we can write down for the divergence of the current, we conclude that
the operator has to have dimension two. Notice that in this case $\tilde j_0$ is parity even and \sayco\ is
consistent with parity. Thus, we could in principle do the same analysis as above. The only point where
something different occurs is at \consi\ where $\tilde j_0$ cannot appear in the right hand side, since it is
inconsistent with the parity of the theory. In addition, all parity odd correlators should be set to zero.
  In this case, parity implies that
\eqn\parti{
[Q, \tilde j_0 ] = \partial^3 \tilde j_0
}
In principle, we have an arbitrary coefficient in this equation but the coefficient is then fixed
by considering various charge conservation identities we mention below.
 When we write the conservation
identity for $\langle j_2 j_2 j_2 \rangle$ we will get the correlator $\la \tilde j_0 j_2 j_2 \rangle $
as the integral term in the right hand side. Again, considering the ward identity on this
last one will require $\langle \tilde j_0 \tilde j_0 j_2 \rangle $ in the right hand side. In this
fashion one can determine the solution. Note that in this case there is only one parameter
which is $\tilde N$.

Interestingly, we can get these correlators by taking the large $\tilde \lambda $ limit of \twopfe \finsfe .
At the level of three point functions of currents the limit is simple to take, namely we see that for any $s \geq 2$
\eqn\limit{
\la s_1 s_2 s_3 \ra \to \la s_1 s_2 s_3 \ra_\bos
}
so that all three point functions become purely boson ones. For $\tilde j_0$, due to \twopfe\ it is necessary
to rescale the operator and define a new operator $  \hat { \tilde j_0 } = \tilde \lambda \tilde j_0 $.
The two point function of $\hat{ \tilde j_0} $ remains finite. This also has the nice feature of
removing $\hat { \tilde j_0}$ from the right hand side of \consi .
The three point functions of the form $\langle \hat {\tilde j_0} j_{s_1} j_{s_2} \rangle $ loose their
$\beta$ structure and remain only with the parity preserving $\gamma_{\tilde 0 s_1 s_2 }$ structure.
%
%
%
The
three point structures involving two scalars survive through a $\beta$ structure, after rescaling the operator. And $\gamma_{\hat 0 \hat 0 \hat 0}$ stays being equal to zero, which is consistent with the large $N$ limit of the
critical $O(N)$ theory. This three point function becomes non-zero at higher orders in the $1/N$ expansion.

Notice that at  $\tilde \lambda =\infty $  the three point functions
 become
parity invariant. However the parity of the operator $\tilde j_0$ got flipped compared to
the one at $\tilde \lambda =0$.



This suggests that the large $\tilde \lambda$ limit
of the fermions plus Chern-Simons matter theory
should agree with the
critical $O(N)$ theory, at least in the large $N$ limit. This was  conjectured in \GiombiKC .
But this conjecture  appears to require a funny relation between the $N$'s and $k's$ of both
theories. In other words, if we start with the quasi-fermion theory with $N$ and $k$ and take
the limit where $\lambda$, defined in  \GiombiKC , goes to one, then the conjecture would
say that this should be the same as the critical $O(N')$ theory with some $k'$ in the limit that
$k' \to \infty$. But the behavior of the free energy in \GiombiKC\ would require that
 \eqn\requ{
 N' \propto N (\lambda -1) [ - \log (\lambda -1 )]^3
 }
 as $\lambda \to 1$. Here the proportionality is just a numerical constant.
 This formula is derived as follows. First notice that \GiombiKC\ derived a formula for the
 free energy, in the large $N$ limit, for a theory of $N$ fermions with a Chern-Simons
 coupling $k$. The conjecture is that this matches a critical bosonic  $O(N')$ theory perturbed by
 a Chern-Simons coupling $k'$. When $\lambda \to 1$ we expect that $k' \to \infty $. So we
 can compute the free energy of the critical bosonic $O(N')$ ignoring the Chern-Simons
 coupling. The free energy of the $O(N')$ theory with no Chern-Simons coupling goes like
 $N'$,   for large $N'$ \SachdevPR .
 Matching the two expressions for the free energy we
 get the relation \requ .
 At first
 sight, \requ\
  seems incompatible with the fact that $N'$ should be an integer. However,
 we should recall that \requ\ is only supposed to be true in the large $N$ (and $N'$) limit.
 Thus, it could be that there is an integer valued function of $N$ and $k$
  which reduces to the right hand side of  \requ\ in the large $N$
 limit.
 Note that if this were true, we would also find the same function in the two point
 function of the stress tensor, the function $f$ discussed in \expo . Of course, here we are
 assuming that our $\tilde \lambda \to \infty $ limit is the same as the $\lambda \to 1$ limit
 in \GiombiKC .

 \subsec{The critical $O(N)$ fermion theory }

Starting from $N$ free fermions, we can add a perturbation $\tilde j_0^2 $. This is
the three dimensional  Gross-Neveu model \GrossJV .
This is an irrelevant perturbation.
And one can wonder whether there is a UV fixed point that leads in the IR to the free fermion plus this
perturbation. In the large $N$ limit, it is easy to see that such a fixed point exists and it is given by
a theory were the operator $\tilde j_0$ in the UV has dimension one \RosensteinPT . Thus it has the properties of the
quasi-boson theory. Again, this is a theory that is parity symmetric.  In \RosensteinPT\  this theory was
argued to be renormalizable to all orders in the $1/N$ expansion.

We can now take the large $g$ limit of the quasi-boson results \twopbo \finsbo .  Again, we need to
rescale $\hat j_0 = \tilde \lambda   j_0$. Only the $\beta_{s_1,s_2,s_3}$ survive in this limit.
With one scalar operator, we get only the $\gamma_{0 s_1 s_2 }$ structure surviving, which is consistent with parity
since now $\hat j_0$ is parity odd.

Notice that,  after the rescaling of the scalar operator,
 $\alpha_{\hat 0\hat 0\hat 0}$ still goes
to zero if we hold $a_3$ fixed.
This is necessary because the three point function of $\la \hat j_0 \hat j_0 \hat j_0 \rangle$
should be zero by parity.

 In principle, we could also introduce, in the large $N$ limit, a
parity breaking interaction of the from $\hat j^3$ which would lead to a non-zero three
point function. This could be obtained by rescaling $a_3$ in \finsbo , so that a finite
term remains.

\ifig\PictureD{The analysis of three point functions can be summarized by
 this picture. The quasi-fermion theory is the top line and the quasi-boson is the bottom line. At the two end points we have the free boson or fermion theory on one side and the
 interacting $O(N)$ theories on the other. This is a statement about the three point functions.
  It would be interesting to understand whether we have a full duality between the two
  theories. Notice that the duality would relate a theory of fermions with a theory with scalars.
  This duality would be a form of bosonization in three dimensions. We also expect
   an RG flow connecting the quasi-boson theory on the top line and the quasi-fermion
   theory on the bottom line for general values of the Chern-Simons coupling. } {\epsfxsize3.0in\epsfbox{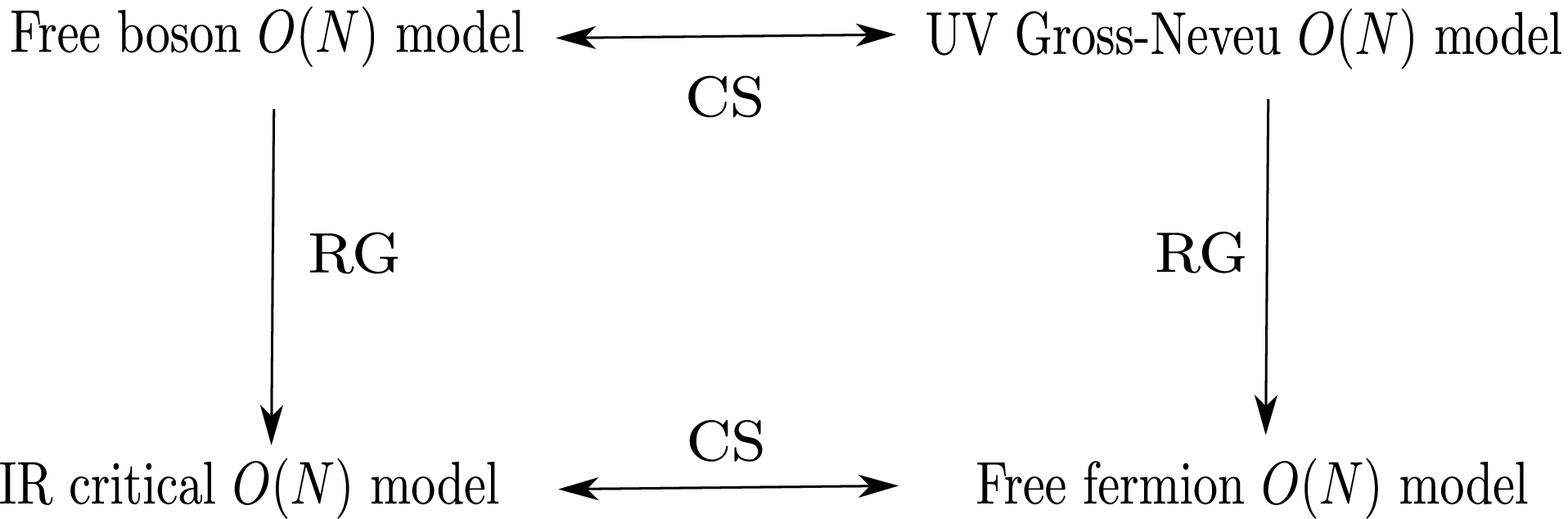}}

\subsec{Relation to  other  Chern-Simons  matter theories }

Thoughout this paper we have focused on theories where the only single trace operators are
given by even spin currents, plus a scalar operator. This is definitely the case for theories
with $N$ bosons or $N$ fermions coupled to an $O(N)$ Chern-Simons theory.

If instead we consider a theory of $N$ complex fermions coupled to a $U(N)$ or $SU(N)$ Chern-Simons
gauge field, then we have additional single trace operators. We still have the same spectrum for
even spins, but we also have additional odd spin currents. However, the theory has a charge
conjugation symmetry under which the odd spin currents get a minus sign. Thus, the odd spin currents cannot
appear in the right hand side of $[Q , j_s]$ where $s$ is even. These odd spin currents can (and do)  appear in
the right hand side of $\nabla .  J_4$ in bilinear combinations. However, since we only considered insertions
of $J_4$ into correlators involving operators with even spin, these extra terms do not contribute.
Thus, the whole analysis in the paper goes through, for these Chern-Simons theories. Our results give the
three point functions of even spin currents\foot{The result for the anomalous
dimension \anmdima\ would be changed by the presence of the odd spin currents.}.
 Presumably a similar analysis can be done for correlators of odd
spin currents, but we will not do this here.  It is also worth mentioning that these
theories {\it do not} have a single trace twist three operator which could appear in the
right hand side of  $\nabla .  J_4 $. Thus the higher spin symmetry breaking only happens
through
double trace operators.



\subsec{Comments about higher dimensions}

We can consider the extension
 of the small breaking ansatz that we explored for $d=3$ to  higher dimensions.
For simplicity we limit ourself to the case of almost conserved currents which are
symmetric traceless  tensors.  We assume that the presence
of a conserved $J_4$ will again fix all three point functions\foot{Though we have not proved this,
the discussion of \MaldacenaJN\ looks almost identical in higher dimensions for symmetric traceless tensors.}. Thus, we are interested in the vector-like scenarios when the conservation of
$J_4$ is broken at ${1 \over N}$.

To analyze this possibility we consider $\pa_{\mu} J^{\mu}_{\ - - -}$ in an  arbitrary number of dimensions. This operator has twist $d$ while
the conserved currents has twist $d-2$. Matching also the spin we get that we can write the following equation
\eqn\divhdim{
\pa_{\mu} J^{\mu}_{\ - - -} = a \pa_{-} {\cal O} J_{--} + b  {\cal O} \pa_{-} J_{--}.
}
where  ${b \over a}$ is fixed by the condition that the right hand side
 is a primary operator.  The scalar operator ${\cal O}$ has
scaling dimension $\Delta = 2$ by matching the quantum numbers.

First, notice that the unitarity bound for the scalar operators is ${d-2 \over 2}$. And, thus, if  we restrict our attention to
unitary theories,  the equation \divhdim\ can be only valid in $d \leq 6$.

We also need to check that there exists a  three point function $\la J_4 J_2 {\cal O} \ra$ that reproduces \divhdim .
Imposing the conservation of $J_2$ leads tho the result \divhdim\ as long
 as $\Delta_{{\cal O}} \neq d-2$. When $d=4$, and $\Delta_{\cal O} =2$,
  the correlator $\la J_4 J_2 { \cal O} \ra $ obeys  $J_4$ current conservation automatically once we impose the $J_2$ current conservation \foot{We thank David Poland for providing us with the Mathematica code to analyze the relevant three point functions in the case of higher dimensions.}.

Thus, we conclude that the scenario that we considered in $d=3$ is impossible to realize in $d=4$\foot{Notice that in $d=4$ we can write also $j_{-} j_{--}$. However, in this case $J_4$ is automatically conserved as soon as we impose conservation of $J_2$ and $J_1$.}. In $d=5$
it seems possible to realize the scenario via the UV fixed point of a  $- (\vec \phi . \vec \phi)^2$ theory. This is a
sick theory because the potential is negative, but one would probably not see the problem
in $1/N$ perturbation theory.  In  $d=6$  we do not know whether there is any example.

\subsec{A comment on higher order  parity violating terms}

 The parity breaking terms in Vasiliev's theories are characterized
by a function $\theta(X) = \theta_0 +   \theta_ 2  X^2 + \cdots $ \Vasiliev .
 The computations we have done
are sensitive to $\theta_0$. At tree level,
the term proportional to $\theta_2$ would start contributing to a
planar five point function, but not to lower point functions \GiombiKC .  However, if we start with
$\theta=0$ and we tried to add $\theta_2$ (keeping $\theta_0=0$),  we run into the
following issues.
For $\theta_0 =0$ we can choose boundary
conditions that preserve the higher spin symmetry. That symmetry fixes all $n$ point
functions \MaldacenaJN . Thus, if turning on $\theta_2$ modifies a five point function, then
it must be breaking the higher spin symmetry. But if we break the higher spin symmetry
we need to modify a three or four point function at the same order.  (To see this, we can just  consider the
correlator of $J_4$ together with the currents that appear in the right hand side of \basicA ).
However, $\theta_2$ would
only contribute to the five point function. Thus, our conclusion is that the function $\theta(X)$
is either constrained to be constant, or it's non-constant part can be removed
by a field redefinition. In this argument we have assumed that these deformations of Vasiliev's
theory in $AdS_4$ also lead to boundary correlators that obey all the properties of a CFT.

\newsec{Conclusions and discussion }

In this paper we have computed the three point correlation functions
 in conformal field theories with a
large $\tN$ approximation, where the higher spin symmetry is broken by $1/\tN$ effects.
Equivalently, we have performed an on shell analysis of Vasiliev theories on $AdS_4$,
constraining the boundary three point correlators. These constraints also apply to the
$dS_4$ case, where they can be viewed as computing possible non-gaussianities in
Vasiliev's theory. From the $dS_4$ point of view we are computing the leading
 non-gaussianities of
the de Sitter  wavefunction\foot{ Of course, for gravitons these match the particular
structures discussed in \MaldacenaNZ , but with particular coefficients. }.

We restricted the theories to contain a single trace  spectrum with only one spin two current
(the stress tensor)  and only one scalar. The scalar can only have dimensions either one or two at leading order in $\tilde N$.
This defines two classes of
correlators which we called quasi-fermion (scalar of dimension two) and quasi-bosons, where the scalar
has dimension one. These are the only possible dimensions that enable the $1/\tN $ breaking of the higher
spin symmetry. The final three point functions depend on an overall constant, $\tN$, which is also the
two point function of the stress tensor. Thus we can view $1/\tN$ as the coupling of the
bulk theory.  In addition, they depend on an extra parameter which selects the
relative weights of the three possible structures in the correlators. The final results are given in
\finsfe , \finsbo . These results apply, in particular, to theories of $N$ bosons or $N$ fermions coupled
to an $SO(N)$ Chern-Simons gauge field. Here $\tN$ scales with $N$ and the extra parameter $\tilde \lambda$ is a
function of $N/k$, where $k$ is the Chern-Simons level, see \expo .

These results also apply to Vasiliev's theories which have parity breaking terms in the
 Lagrangian \Vasiliev . It also
applies to Vasiliev's theories with boundary conditions that break the higher spin symmetry, but preserve
conformal invariance.

At strong coupling the quasi-fermion three point functions go into the three point functions of the
critical $O(N)$ theory. This suggests that there should be a duality between the small $k$ limit of the
Chern-Simons theories and the critical $O(N)$ theory  \GiombiKC . However,
 the thermal partition function computed in \GiombiKC\ suggests that if this duality is true, the
 connection between the values of $N$ and $k$ of the two theories is rather intricate, see \requ .
It would be nice to further understand  this issue, and to find the correct duality, if there is one.
Naively, { \it if } our methods actually fix all $n$ point functions, then all leading order
correlators would be consistent with the duality. This duality would be
a three dimensional version of bosonization.

Our method was based on starting with
  the simplest possible even spin higher spin current, $J_4$, and writing the most general form for its
   divergence. We then noted that the violation of conservation of the $J_4$ current leads to
   constraints on three point functions, which we solved. These equations were  mostly
   identical to the
   ones one would get in the conserved case, except when acting on correlators with spins two or lower. In these
   cases we obtained some relations which eliminated some of the free parameters and left only two parameters
   in the quasi-fermion theory and three parameters in the quasi-boson case. These parameters match with   the
   free parameters that we have in large $N$ Chern-Simons models.

Since we only considered the current $J_4$ one can wonder whether it is possible to add other deformation parameters
which affect current conservation for higher spin currents but not $J_4$.
 We argue in appendix F that this
is not possible.

An interesting extension of these results would be to carry out this procedure for higher point functions. This
is in principle conceptually straightforward, but it seems computationally difficult.
Note that this analysis amounts to an on-shell study of the Vasiliev theory. We study the
physical, gauge invariant, observables of this theory with $AdS_4$ or $dS_4$ boundary conditions.
As it has often been emphasized, the on shell analysis of gauge theories can be simpler
than doing computations in a fully Lorentz invariant formulation.

 The methods discussed in this paper could be viewed as  on-shell methods to compute correlation functions
   in certain matter plus Chern-Simons theories. Note that we did not have any gauge fixing issues, since we never considered gauge non-invariant quantities. These methods apply only to the special class of theories
   that do not contain a twist three, spin three single trace operator that can directly Higgs the $J_4 $ higher
   spin symmetry already at leading order in $N$. It would also be interesting to consider cases
   where this Higgsing can happen already for single trace operators.
   If the mixing with this other operator is small, which occurs in weakly coupled theories, we can probably
   generalize the discussion of this paper. We would only need to add the twist three single trace operator in
   the right hand side of the divergence of the current,  $\nabla .  J_4$. This would lead to an on shell method for
computing correlators. Something in this spirit was
 discussed  in \AnselmiMS ,  \HennMW . This Higgsing mechanism, named ``La Grande Bouffe'' in \BianchiWX\  is also important for understanding the emergence of a more ordinary looking
 string theory in AdS from the higher spin system.

Our analysis also works for  higher spin theories
 on de-Sitter space. In that case, we are  constraining the wavefunction of the universe.
 It is valid for the proposed examples of dS-CFT \refs{\AnninosUI, \Ng}, as well as further
 examples that one might propose by looking at the parity violating versions of the Vasiliev
 theory. To consider these cases one should set $\tilde N < 0 $ in our formulas . Of course, the constraints hold whether we know the CFT dual or not!

 We have restricted our analysis to the case of $AdS_4/CFT_3$. Recently,  examples of
 $AdS_3/CFT_2$ theories with higher spin symmetry have been considered. See for example
  \refs{\GaberdielPZ, \GaberdielWB, \GaberdielZW, \ChangMZ, \PapadodimasPF, \ChangVK}.
   In lower dimensions the higher spin symmetry appears less restrictive, so one would need a
 more sophisticated analysis than the one presented in this paper.
  On the other hand, in higher dimensions the higher spin symmetry is more constraining. The
  higher dimensional case will hopefully  be discussed separately.

We have used conformal symmetry in an important way, it would also be interesting
to study non-conformal cases. For example, we can expect to constrain the flows between
the top and bottom lines of figure 3.

{\bf Acknowledgments}

  We would like to thank N. Arkani-Hamed, D. Gross,
  S. Giombi,  T. Hartman, K.  Hinterbichler,  D. Poland, S. Rychkov,
 N. Seiberg, P. Sundell,  X. Yin and   E. Witten for discussions.

This work   was  supported in part by U.S.~Department of Energy
grant \#DE-FG02-90ER40542. The work of AZ was supported in part by the US National Science Foundation under Grant No.~PHY-0756966 and by the Department of Energy under Grant No.\#DE-FG02-91ER40671.

\appendix{A}{Why the divergence of the currents  should be conformal primaries}

In this appendix we show  that the divergence of an operator with  spin becomes a
conformal  primary
when $\tau - 1 \to 0$. What happens is that the large representation with twist $\tau$ and
spin $s$ is splitting into a representation with $\tau =1$ and spin $s$ plus a representation
with $\tau =3$ and spin $s-1$. This second representation is what appears in the right hand
side of the divergence of $J_s$ as $\tau \to 1$. This is well known fact, that was used
in the past in \refs{\HeidenreichXI,\AnselmiMS,\HennMW,\GirardelloPP}.
Here we recall its proof for completeness.

Let us normalize the current to one,  $ \la J_s | J_s \ra $. Let us
define
\eqn\divod{
 \partial . J_s = \alpha O_{s-1}
 }
  where $O_{s-1}$ is also normalized to one. Then we have
\eqn\derivannih{\eqalign{
\la \pa J | \pa J \ra &\propto (\tau - 1) \la J | J \ra = \tau - 1  \cr
\la \pa J | \pa J \ra  &= \alpha^2
}}
Where in the first line we converted  the derivative in the bra into
a special conformal generator in the ket, and then commuted it through the derivative in the ket
 using the conformal algebra. We ignored numerical factors.
  In the  second line we used \divod .   The conclusion is that
 \eqn\sizeal{
 \tau -1 \propto \alpha^2
 }
We now act on both sides of \divod\ with the special conformal generator $K_\nu$. On
the left we use the conformal algebra to evaluate the answer.
We then get that $\alpha K_\mu O_{s-1}  \propto (\tau -1) J_\mu \propto \alpha^2 J_\mu $.
The conclusion is that
$ \la K_\mu O_{s-1} | K_\mu O_{s-1} \ra \propto \alpha^2 $. Thus we see that, to leading
order in $\alpha$,  the operator  $O_{s-1}$ is a conformal primary.
Of course this is true regardless of whether $O_{s-1}$ is a single trace or multitrace operator,
as long as $\tau - 1 $ is very small.

Finally, note that if we know $\alpha$ appearing in \divod\ then we can compute the
anomalous dimension $\tau -1$ of the current via the same formulas.

In particular, in the quasi-fermion theory we have argued around \fins\ that
$a_2\sim \tilde \lambda/ \tN $. This together with the value of the $\tilde j_0$ two
point function \normtwo\ implies that for the spin four current we have
\eqn\anmdi{
\tau_4 -1  = { 32 \over 21 \pi^2 } { 1 \over \tilde N }  {\tilde  \lambda^2 \over 1 + \tilde \lambda^2  }
}
This formula should be applied to $O(N)$ theory.

For general group and arbitrary spin $s$ we expect the following formula to be true
\eqn\anmdigen{
\tau_s -1  = a_s { 1 \over \tilde N }  {\tilde  \lambda^2 \over 1 + \tilde \lambda^2  }  + b_s { 1 \over \tilde N }  {\tilde  \lambda^2 \over (1 + \tilde \lambda^2 )^2 }
}
where $a_s$ and $b_s$ are some fixed numbers.

In principle these arguments allow us to fix the overall numerical coefficient.
However, as a shortcut, we have
used the  formula for the anomalous dimensions  for the critical
$O(N)$ theory given in eqn. (2.20) of \RuhlBW . Thus, we fixed the overall coefficient
in \anmdi\ so that the $\tilde \lambda \to \infty$ limit matches \RuhlBW . This also
fixes $a_s = { 16 \over 3 \pi^2 } { s-2 \over 2 s -1} $.

\appendix{B}{Structure of the pseudo-charge conservation identities }

In this appendix we recall how the various coefficients that appear in the charge conservation
identity are fixed. Assume that we consider only twist one three point functions, and that the spins are all
non-zero. Then we can use \normal , perhaps including a $j_0^\eff$ operator when it acts on $j_2$.
Then the action of $Q$ on $\langle j_{s_1} j_{s_2} j_{s_3} \rangle$ gives an expression of the form
\eqn\expresfu{
 \left[  \sum_{i=0,\pm 2 } r_{\bos, 1,i}  \partial^{3 - i } \langle j_{s_1 + i } j_{s_2} j_{s_3} \rangle_\bos  +
   {\rm cyclic }\right]  + [ \bos \to \fer ] + [ \bos \to \odd ]
}
Here we have the set of coefficients $r_{{\rm type} ,a , i}$. Here type goes over $\bos$, $\fer$, $\odd$.
$a$ labels the point and runs over 1,2,3. Finally $i$ runs over $\pm 2, 0$. In total there are up to
 $3^3 = 27$ coefficients.
These coefficients result from multiplying the $c_{s,s'}$ in \normal\ and the $\alpha$, $\beta$ and $\gamma$ in
\writen .

These equations split into three sets of equations, one for each type.
In each set of equations the coefficients in front of different three point functions  are all fixed up to an overall constant, except in the cases
where the corresponding three point structure vanishes automatically, where, obviously, the corresponding
coefficient is not fixed.

For the boson and fermion types this follows from the discussion in appendix J of \MaldacenaJN .
One simply needs to take successive light-cone limits of the three possible pairs of particles to
argue that the coefficients are all uniquely fixed.

For the odd structure the situation is more subtle since some of the equations do not have any non-zero
solutions if we restrict to structures inside the triangle rule $s_i \leq  s_{i+1} + s_{i-1} $. However, there
are non-zero solutions once we take into account that the current non-conservation allows solutions outside the
triangle. We have checked this in some cases, and we think it is likely to be true in general, but we did not
prove it. We know that there is at least one solution with non-zero coefficients, which is the
one that the Chern-Simons construction would produce. Thus, in order to show that the solution is unique, we would need to show that there is no solution after we set one of the coefficients to zero. We leave
this problem for the future.
\appendix{C}{Compendium of normalizations}

Here we would like  to present more details
 on the normalization convention that we chose for the currents.
 Noticed that we have set them by the choice \normco . Here we will check that with \normco\
 the coefficients of all the terms in \normal\ are the same for the single free boson and
 single free fermion theory. This is related to the fact that the higher spin algebra is the same
 for bosons and fermions in three dimensions (see, for example, \SezginPT).

In this appendix we do computations in the free theories. Then it is convenient to consider
the matrix elements of the currents in Fourier space.
As
explained
in the Appendix J of \MaldacenaJN\ it is convenient to introduce a combination of the
two momenta that appear in the on shell matrix elements of a current with two on shell fermions or
bosons . The idea is roughly to change $\pa^m \psi^{1}_{-} \to (z+\bar{z})^{2 m+1}$,
$\pa^k \psi^{2}_{-} \to (-1)^k (z-\bar{z})^{2 k+1}$. Where the indices 1 and 2 denote the two fermion fields that make up the current.
Then currents take a form
\eqn\curr{
j_{s} = \alpha_s \left[ z^{2 s} - \bar{z}^{2 s} \right]
}
where we introduced normalization factor $\alpha_s$ explicitly.

The charge generated by $j_4$ that we consider in the main text is given then by
\eqn\charge{
Q = \alpha_{Q} \left[ (z + \bar{z})^{6} - (z - \bar{z})^{6} \right]
 \propto \left[ \partial_{x^-_1}^3 + \partial_{x^-_2 } ^3 \right]
}
where we again introduced the normalization factor for the charge. We have written the
action of the charge on the two free fields that make the current.
 Also notice that $\alpha_2$ is fixed in a canonical way.

Now we can start fixing the normalizations. We start from $c_{2,4} = 1$. Using $[Q, j_2]$ we get
\eqn\ctwofour{
\alpha_{Q} = {\alpha_4 \over 12 \alpha_2 }
}
then we consider $[Q, j_4]$ to fix $c_{4,4} = 1$, $c_{4,6} = 1$. It gives
\eqn\nextstepc{\eqalign{
\alpha_4 &= {3 \over 10} \alpha_2 \cr
\alpha_6 &= \left( {3 \over 10} \right)^2 \alpha_2
}}
simple analysis further shows that
\eqn\nextstepB{
\alpha_{s} = \left( {3 \over 10} \right) \alpha_{s-2}
}
this follows from the formula (J.12) in \MaldacenaJN\ which can be  written as
 \eqn\tocheck{\eqalign{
[Q_s, j_{s'}]  = &  ( 2 s-2 )! \sum_{r = 0}^{s+s'-2} \tilde \alpha_r (s,s') \partial ^{ s+s'-r -1}j_{r}
\cr
 \tilde \alpha_r  (s,s') = &  \left[1 + (-1)^{s+s'+r} \right]  \left( { 1 \over \Gamma( r+ s - s') \Gamma(s+s'  -r) } \pm { 1 \over \Gamma(r  + s + s') \Gamma(s-s' -r) }\right).
 }}
Notice that in this normalization $c_{2,2}$ is also fixed to one automatically. Here the
$\pm$ sign corresponds to the boson or fermion case.

The case of bosons  is almost identical.
It is important that that the term with $\pm$ in \tocheck\ is zero for all terms $
\tilde \alpha_{r}(4,s')$ with $r, s' \geq 2$. These are all the cases where we can compare
the normalization between fermions and bosons.  This implies that all constants
$c_{s,s}$ and $c_{s,s \pm 2 }$ that appear in \normal , are the same in the free boson or
free fermion theories.
%
%
%
The case when one of the spins is zero, corresponds to the appearance of the scalar operator
of twist one $j_0$ which is only present in the free boson theory.

Moreover, using the formulas above and the results of \AnselmiBB\ one can check that in the normalization we are using $n_{s}^{\rm free \ boson} = n_{s}^{\rm free \ fermion}$. More precisely, one can first relate the normalization used in \AnselmiBB\ to the normalization here and then used the results of \AnselmiBB\ for two point functions to get that
\eqn\restwop{\eqalign{
n_{s}^{\rm free \ boson} &= n_{s}^{\rm free \ fermion} = \alpha_s^2 \Gamma (1 + 2 s) \cr
\alpha_2 &= {1 \over 16 \pi}, ~~~ \alpha_{s} = ({3 \over 10}) \alpha_{s-2}.
}}
We used this result in the main text.

After we fix all normalizations of currents in this way we can compute all three point functions in the free fermion theory $\la j_{s_1} j_{s_2} j_{s_3} \ra_{\fer }$ (as well as in the theory of free boson to get $\la j_{s_1} j_{s_2} j_{s_3} \ra_{\bos }$). These are the solutions that we use in the main text,
 for example, in \writen  .

\appendix{D}{Exploring the scalar sector for the quasi-fermion}

\def\cp{{ \pa^{2\perp} }}
\def\cu{{\pa^3 }}

In this appendix we explain how the scalar sector correlation functions could be recovered. We start from writing the general form
of the variation of the scalar operator $\tilde j_0$ under the action of $Q$
\eqn\scalaraction{\eqalign{
[Q, \tilde j_0] = & \tilde c_{\tilde 0 , \tilde 0}  c_{\tilde 0 , \tilde 0} \pa^3 \tilde j_{0} + \tilde c_{\tilde 0 , 2}  c_{\tilde 0 , 2} \pa \left( \pa_{-} J_{y-} - \pa_{y} J_{--} \right)
\cr
[ Q, \tilde 0 ] =&  \tilde c_{\tilde 0 , \tilde 0}  c_{\tilde 0 , \tilde 0}  \pa^3 \tilde {0} + \tilde c_{\tilde 0 , 2}  c_{\tilde 0 , 2}  \cp
 2
}}
where remember coefficients $c$ are the ones we would get in the theory of free fermion with the normalization of the operators that we chose.
The interesting part is a deviation from the free theory which we denoted by $\tilde c$ following the notations used in the main text. On the second line we introduce a shortened notation where
we replaced $j_s \to s $ and in addition, we also introduced the symbol $\cp$ to denote
the combination of derivatives and indices of $J_2$ that appear in the first line.

To fix some of the coefficients we use the $\la 4 2 \tilde 0 \ra$ three point function.
 It has two different structures: fermion and odd ones and we have already determined their
 coefficients.
It is easy to see that it leads to the identities
\eqn\WIvarfact{\eqalign{
\gamma_{4 2 \tilde 0} &\propto \tilde \lambda n_{\tilde 0} \cr
 \tilde c_{\tilde 0 , 2}  c_{\tilde 0, 2} n_2 &=  \beta_{4 2 \tilde 0 }
}}
In the first line we took the divergence of the $J_4$ current and compared to \sayco .
In the second line we integrated the current to get the charge acting on $\tilde j_0$.
From the first line we get that $\tilde n_{\tilde 0} = {1 \over 1 + \tilde \lambda^2}$ where we used the fact that at $\tilde \lambda = 0$ it should be equal to $1$. The second equation then fixes
\eqn\WIsecfixvar{
\tilde c_{\tilde 0 , 2} = {1 \over 1 + \tilde \lambda^2}.
}

\subsec{Fixing $\tilde c_{\tilde 0 , \tilde 0}$ and nontrivial consistency check}

Here we fix $\tilde c_{\tilde 0 , \tilde 0}$ by considering the pseudo-conservation
identities
for $\la \tilde 0 s_1 s_2 \ra$ where both $s_1$ and $s_2$ are larger than $2$ for simplicity.
We will encounter a rather intricate structure for this identity.

We schematically write relevant terms in pseudo-conservation identities with their $\tilde \lambda$ scaling
\eqn\puzzlingWI{\eqalign{
&\tilde c_{\tilde 0 , \tilde 0}  {1 \over 1+\tilde \lambda^2}
\la  \cu  \tilde 0 s_1 s_2 \ra_{\fer} + \tilde c_{\tilde 0 , \tilde 0}  {\tilde \lambda \over 1+\tilde \lambda^2} \la  \cu \tilde 0 s_1 s_2 \ra_{\odd} + \cr
& {1 \over 1+\tilde \lambda^2} \left( {1 \over 1+\tilde \lambda^2}  \la \cp  2 s_1 s_2 \ra_{\fer} + {\tilde \lambda \over 1+\tilde \lambda^2}  \la  \cp 2 s_1 s_2 \ra_{\odd} + {\tilde \lambda^2 \over 1+\tilde \lambda^2}  \la  \cp 2 s_1 s_2 \ra_{\bos}  \right) + \cr
 & {1 \over 1 + \tilde \lambda^2} \la \tilde 0 \ {\rm standard \ terms}\ra_{\fer} + {\tilde \lambda \over 1 + \tilde \lambda^2} \la \tilde 0 \ {\rm standard \ terms}\ra_{\odd} = \cr
& ={\tilde \lambda \over 1 + \tilde \lambda^2} \partial_{x_1^-} \int {1 \over |x - x_1|^4}  \left( {1 \over 1+\tilde \lambda^2}  \la  2 s_1 s_2 \ra_{\fer} + {\tilde \lambda \over 1+\tilde \lambda^2}  \la 2 s_1 s_2 \ra_{\odd} + {\tilde \lambda^2 \over 1+\tilde \lambda^2}  \la  2 s_1 s_2 \ra_{\bos}  \right)
}}
Here, except for the terms involving $\cp 2$ all components of the currents are minus, as are all
derivatives.
Now let's start looking for a  solution. All $\tilde \lambda $ dependence is explicit in this
equation.
Note that the equations contain terms with various $\tilde \lambda$ dependence.
First we match the double poles at $\tilde \lambda^2 = -1$.
This requires the following two identities
\eqn\substr{\eqalign{
\la \cp 2 s_1 s_2 \ra_{\odd} &= \partial_{x_1^-}  \int {1 \over |x - x_1|^4} \left( \la  2 s_1 s_2 \ra_{\fer} - \la  2 s_1 s_2 \ra_{\bos} \right),  \cr
\partial_{x_1^-} \int {1 \over |x - x_1|^4}  \la 2 s_1 s_2 \ra_{\odd} &=
- \la \cp  2 s_1 s_2 \ra_{\fer} + \la \cp  2 s_1 s_2 \ra_{\bos}.
}}\
The second  line can also  be used to make all terms of the $_\fer $ terms work. This
would work nicely if we set $\tilde c_{\tilde 0 , \tilde 0 } =1$, which is what we wanted to
argue.  In fact, replacing the second line, we get all the parity odd pieces in the full equation
work out. By parity odd we mean the terms that
are odd  under $y \to -y$.
%
%

We are then left with all terms that are even under $y \to - y$.
After using the first equation in \substr\ we are left with the condition
\eqn\canceloddD{\eqalign{
&\la  \cu \tilde 0 s_1 s_2 \ra_{\odd}+ \la \tilde 0 \ {\rm standard \ terms}\ra_{\odd} = \partial_{x_1^-}
\int {1 \over |x - x_1|^4}  \la  2 s_1 s_2 \ra_{\bos}
}}

To summarize,  in this subsection we fixed $\tilde c_{\tilde 0 , \tilde 0} = 1$ and presented
 the self-consistency relations \substr \canceloddD\
 that are necessary for the whole construction to work. We checked one of the \substr\ in the light cone limit and it indeed works.  It would be nice to check these identities more fully.

\subsec{Fixing $\la s \tilde 0 \tilde 0 \ra$}

To fix these three point functions we consider pseudo-conservation identities that we get from $\la s \tilde 0 \tilde 0 \ra$. Notice that we already know that $\la 2 \tilde 0 \tilde 0 \ra$ and $\la 4 \tilde 0 \tilde 0 \ra$ functions are given by $\beta_{222}$.  The first follows from the stress tensor
Ward identity and the two point function. The second
 follows from the action of the charge $Q$ on $\tilde j_0$, with $\tilde c_{\tilde 0 , \tilde 0 } =1$
 that we derived above.

Starting from this we can  build the induction. Consider first pseudo-conservation  identity for $\la 4 \tilde 0 \tilde 0 \ra$. We get schematically
\eqn\terms{\eqalign{
&\beta_{6 \tilde 0 \tilde 0} \la \partial 6 \tilde 0 \tilde 0\ra_\fer
+ {1 \over {1+ \tilde \lambda^2}} \left(\la \partial^5 2 \tilde 0 \tilde 0\ra_\fer
+ \la \partial^3 4 \tilde 0 \tilde 0\ra_\fer  \right) +  \cr
&{1 \over 1 + \tilde \lambda^2} \left( {1 \over 1 + \tilde \lambda^2}  \la 4 \cp
2 \tilde 0\ra_\fer +  {\tilde \lambda \over 1 + \tilde \lambda^2}  \la 4\cp  2 \tilde 0\ra_\odd +    \la 4 \partial^3\tilde 0 \tilde 0\ra_\fer  + \tilde \lambda \la 4 \partial^3 \tilde 0 \tilde 0\ra_\odd  \right) +  x_2 \leftrightarrow x_3 \cr
&= {\tilde \lambda \over 1 + \tilde \lambda^2}
\partial_{x_2^-} \int {1 \over |x - x_1|^4}  \left( {1 \over 1 + \tilde \lambda^2}  \la 4 2 \tilde 0\ra_\fer
+  {\tilde \lambda \over 1 + \tilde \lambda^2}  \la 4 2 \tilde 0\ra_\odd  +
[ x_2 \leftrightarrow x_3]  \right),
}}
%
Again, matching the double poles at $\tilde \lambda^2 =-1$ requires
%
\eqn\twoneweq{ \eqalign{
\la 4 \cp 2 \tilde 0\ra_\odd  = &   \partial_{x_2^-} \int {1 \over |x - x_1|^4} \la 4 2 \tilde 0\ra_\fer,
\cr
\la 4 \cp 2 \tilde 0\ra_\fer  = &   - \partial_{x_2^-} \int {1 \over |x - x_1|^4} \la 4 2 \tilde 0\ra_\odd .
}}
Then replacing these identities in the equations implies that the odd piece cancels and
the fermion piece works if
%
%
%
\eqn\generalanswer{
\beta_{6 \tilde 0 \tilde 0} = \beta_{2 2 2} = { \tN  \over 1 + \tilde \lambda^2 } .
}
Then we proceed by induction. Considering pseudo-charge conservation identities for $\la s \tilde 0 \tilde 0 \ra$ we get a similar story. Double pole matching leads to
\eqn\twoneweqGEN{ \eqalign{
\la s \cp 2 \tilde 0\ra_\odd  = &   \partial_{x_2^-} \int {1 \over |x - x_1|^4} \la s 2 \tilde 0\ra_\fer,
\cr
\la s \cp 2 \tilde 0\ra_\fer  = &   - \partial_{x_2^-} \int {1 \over |x - x_1|^4} \la s 2 \tilde 0\ra_\odd ,
}}
and the fermion piece fixes
\eqn\generalanswerGEN{
\beta_{s \tilde 0 \tilde 0} = \beta_{2 2 2} = { \tN  \over 1 + \tilde \lambda^2 } .
}

This fixes the three point function involving two $\tilde j_0$ operators.

\subsec{Fixing $\gamma_{\tilde 0 \tilde 0 \tilde 0 }$}

To fix the last three point function we consider the WI $\la \tilde 0 \tilde 0 \tilde 0 \ra$. We get schematically the equation
\eqn\lastequation{
\gamma_{\tilde 0 \tilde 0 \tilde 0}(\pa^3_1 + \pa^3_2 + \pa^3_3 ) \la \tilde 0 \tilde 0 \tilde 0 \ra_\odd
 = { \tilde \lambda n_0 \over \tilde N} \beta_{2 \tilde 0 \tilde 0} \sum_{perm} \pa_{x_1^{-} }\int {1 \over |x - x_1|^4} \la 2 \tilde 0 \tilde 0 \ra_{\fer}.
}

The right hand side of this
equation can be computed explicitly using the star-triangle identity (see, for example, \KazakovKM).   More precisely,
 each term in the right hand side of \lastequation\
  can be rewritten  (thanks to our sandwich geometry as we can pull out the derivatives)  as
\eqn\rewriteX{
{1 \over x_{23}^{3/2}}\pa_{x^-_1} \left(\pa_{x^-_2}^2 + \pa_{x^-_3}^2 - 6 \pa_{x^-_2} \pa_{x^-_3} \right) \int d^3 x {1 \over |x - x_1|^2 |x - x_2|^{3} |x - x_3|} + {\rm perm.}
}
The integral is finite and is
given by ${ x_{23} \over x_{12}^2 x_{13}^2}$. By taking the derivatives and summing over permutations which come from different contractions in \lastequation\ one can check
 that the sum is actually zero. Thus we are forced to set
\eqn\lastsectorX{
\gamma_{\tilde 0 \tilde 0 \tilde 0} = 0.
}
This is nicely consisted with the critical $O(N)$ limit.

\appendix{E}{Exploring the scalar sector for the quasi-boson}

Now we would like to go through the same analysis but for the quasi-boson theory.

We start by
 writing the general form of the variation of the scalar operator $j_0$ under the action of $Q$
\eqn\scalaraction{
[Q,  j_0] =\tilde c_{ 0 ,  0}  c_{0 , \tilde 0} \pa^3 j_{0} + \tilde c_{0 , 2}  c_{0 , 2} \pa j_{2}
}
Recall that  coefficients $c$ are the ones we would get in the theory of free boson with the normalization of the operators that we chose.
The interesting part is a deviation from the free theory
which we denoted by $\tilde c$ following the notations used in the main text.

First we consider the $\la 4 2 0 \ra$ three point function. It has two different structures:  boson and odd ones.
This leads to the identities
\eqn\WIvarfact{\eqalign{
\gamma_{4 2 0} &\propto \tilde \lambda n_{0} \cr
 \tilde c_{0 , 2}  c_{ 0,  2} n_2 &= \alpha_{420}= \alpha_{222}
}}
In the first line we simply have taken the divergence of $J_4$ and used \divscalarNice .
In the second line we have integrated $J_4$ around $j_0$.
From the first line we get that $\tilde n_{0} = {1 \over 1 + \tilde \lambda^2}$ where we used the fact that at $\tilde \lambda = 0$ it should be equal to $1$. The second equation then fixes
\eqn\WIsecfixvar{
\tilde c_{0 , 2} = {1 \over 1 + \tilde \lambda^2}.
}

\subsec{Fixing $\tilde c_{0 , 0}$}

To fix $\tilde c_{0 , 0}$ we are analogously considering the pseudo-conservation identities
for $\la 0 s_1 s_2 \ra$ where both $s_1$ and $s_2$ are larger than $2$ for simplicity.

We schematically write relevant terms in pseudo-conservation identities with their $\tilde \lambda$ scaling
\eqn\puzzlingWI{\eqalign{
&\tilde c_{ 0 , 0}  {1 \over 1+\tilde \lambda^2} \la  \cu 0 s_1 s_2 \ra_{\bos} + \tilde c_{0 , 0}  {\tilde \lambda \over 1+\tilde \lambda^2} \la \cu 0 s_1 s_2 \ra_{\odd}+ \cr
& {1 \over 1+\tilde \lambda^2} \left( {1 \over 1+\tilde \lambda^2}  \la \pa  2 s_1 s_2 \ra_{\bos} + {\tilde \lambda \over 1+\tilde \lambda^2}  \la \pa 2 s_1 s_2 \ra_{\odd} + {\tilde \lambda^2 \over 1+\tilde \lambda^2}  \la  \pa 2 s_1 s_2 \ra_{\fer}  \right) + \cr
& {1 \over 1 + \tilde \lambda^2} \la 0 \ {\rm standard \ terms}\ra_{\bos} + {\tilde \lambda \over 1 + \tilde \lambda^2} \la 0 \ {\rm standard \ terms}\ra_{\odd} = \cr
& ={\tilde \lambda \over( 1 + \tilde \lambda^2)^2 } \int {1 \over |x - x_1|^2}  \left(
 \la  \cp 2 s_1 s_2 \ra_{\bos} +
 \tilde \lambda
 \la \cp  2 s_1 s_2 \ra_{\odd} +
\tilde \lambda^2  \la  \cp 2 s_1 s_2 \ra_{\fer }  \right)
}}
Again we match the double poles at $\tilde \lambda^2 = -1$ to obtain
\eqn\idencovu{ \eqalign{
 \la \pa 2 s_1 s_2 \ra_{\odd}
& = \int {1 \over |x - x_1|^2}  \left(  \la  \cp 2 s_1 s_2 \ra_{\bos} -    \la  \cp 2 s_1 s_2 \ra_{\fer }  \right)
\cr
-  \la \pa  2 s_1 s_2 \ra_{\bos} +  \la  \pa 2 s_1 s_2 \ra_{\fer}
& = \int {1 \over |x - x_1|^2}    \la \cp  2 s_1 s_2 \ra_{\odd}
}}
Now using these equations, we find that the all the even pieces under $y \to -y$ work
properly only if
  $\tilde c_{0,0} =1$.
Then the odd piece reduces to the condition
\eqn\cancelodd{\eqalign{
&  \la  \cu 0 s_1 s_2 \ra_{\odd} + \la  0 \ {\rm standard \ terms}\ra_{\odd}  =
\int {1 \over |x - x_1|^2}   \la  \cp 2 s_1 s_2 \ra_{\fer}
}}
We have not checked explicitly whether \cancelodd\ and \idencovu\ are true, but they should be for
consistency.

\subsec{Fixing $\la s 0 0 \ra$}

To fix these three point functions we consider pseudo-conservation identities that we get from $\la s 0 0 \ra$. Notice that we already know that $\la 2  0  0 \ra$ and $\la 4  0 0 \ra$ functions are given by $\alpha_{222}$. This follows from the stress tensor Ward identity and from the action of
$Q$ on $j_0$, together with the normalization of $j_0$.

Starting from this we can  build the induction. Consider first pseudo-conservation  identity for $\la 4 0 0 \ra$. We get schematically
\eqn\terms{\eqalign{
&\alpha_{ 6 0 0} \la \pa 6 0 0\ra_\bos + {1 \over {1+ \tilde \lambda^2}} \left(\la \pa^5 2 0 0\ra_\bos + \la \pa^3 4 0 0\ra_\bos \right) + \cr
&+ {1 \over 1 + \tilde \lambda^2} \left( {1 \over 1 + \tilde \lambda^2}  \la 4 \pa 2 0\ra_\bos +  {\tilde \lambda \over 1 + \tilde \lambda^2}  \la 4 \pa 2  0\ra_{\odd} \right) +
\cr
&+ {1 \over 1 + \tilde \lambda^2}  \la 4 \pa^3 0 0\ra_\bos +  {\tilde \lambda \over 1 + \tilde \lambda^2}  \la 4 \pa^3 0  0\ra_{\odd} + [ x_2 \leftrightarrow x_3 ]= \cr
&= {\tilde \lambda \over 1 + \tilde \lambda^2} \int {1\over |x - x_1|^2}  \left( {1 \over 1 + \tilde \lambda^2}  \la 4   \cp 2 0\ra_\bos +  {\tilde \lambda \over 1 + \tilde \lambda^2}  \la 4 \cp 2  0\ra_{\odd} + [ x_2 \leftrightarrow x_3 ] \right),
}}

Again looking at the double pole at $ \tilde \lambda^2 =-1$ we get the equations
\eqn\newidne{ \eqalign{
 \la 4 \pa 2 0\ra_\bos = & -
\int {1\over |x - x_1|^2}    \la 4 \cp 2  0\ra_{\odd}  ,
\cr
 \la 4 \pa 2 0\ra_\odd = &
\int {1\over |x - x_1|^2}    \la 4 \cp 2  0\ra_{\bos}
}}
This implies that
$\alpha_{6 0 0} = \alpha_{222}$.

By induction through the identities
\eqn\newidnegs{ \eqalign{
 \la s \pa 2 0\ra_\bos = & -
\int {1\over |x - x_1|^2}    \la s \cp 2   0\ra_{\odd}  ,
\cr
 \la s \pa 2 0\ra_\odd = &
\int {1\over |x - x_1|^2}    \la s \cp 2  0\ra_{\bos}
}}
we get that
\eqn\generalanswer{
\alpha_{s 0 0} = \alpha_{2 2 2}.
}

Again, we did not verify the identities \newidnegs\ but they are necessary for consistency.

\subsec{Vanishing of the double trace term  in the $\la 000 \ra$ identity}

When we studied the pseudo-charge conservation identity for the triple scalar correlator
$\la j_0 j_0 j_0 \ra $ in the main text we said that the double trace term in the
divergence of $J^4$, \divscalarNice , vanishes. Here we prove that assertion.

We start from the following integral
\eqn\estartxxx{
I(x_1 , x_2 , x_3) = a_2 \int d^3 x \la j_0 (x) j_{0} (x_1) \ra \la \cp J_2 (x) j_0 (x_2) j_0 (x_3) \ra
}
where operator $j_0(x) \cp J_2(x)$ comes from  an insertion of the double trace term in
 \divscalarNice .
After some algebra one can re-express this as the following integral
\eqn\estartxxxB{
I(x_1 , x_2 , x_3) \sim  \pa_{x_1^{-}}  \left( \pa_{x_2^{-}} -  \pa_{x_3^{-}} \right) \left( \pa_{x_2^{-}} \pa_{y_3}-  \pa_{x_3^{-}} \pa_{y_2} \right)  \int d^3 x {1 \over |x - x_1|^2 |x - x_2| |x - x_3|}
}
written in this way it is manifestly finite. However, we rewrite it again as follows
\eqn\estartxxxC{
I(x_1 , x_2 , x_3) \sim  \pa_{x_1^{-}}  \left( \pa_{x_2^{-}} -  \pa_{x_3^{-}} \right) \left[ y_{2 3} \pa_{x_3^{-}} J(x_1,x_2, x_3) - {1 \over 2} x_{2 3}^{+} \pa_{y_3} J(x_1,x_2, x_3) \right]
}
where
\eqn\jintdef{
J(x_1, x_2, x_3) = \int d^3 x {1 \over |x - x_1|^2 |x - x_2|^{3} |x - x_3|}
}
is the conformally invariant integral. It is divergent, however, the difference in \estartxxxC\ is, of course, finite. We can take this integral using the well-known star-triangle identity. We regularize the integral as
\eqn\jintdefreg{
J_{\delta }(x_1, x_2, x_3) = \int d^{3+\delta} x {1 \over |x - x_1|^{2+\delta} |x - x_2|^{3} |x - x_3|}
}
using the star-triangle formula, expanding in $\delta$ and plugging into \estartxxxB\ we see that the divergent piece cancels and the finite piece is given by
\eqn\eeestartD{
I(x_1 , x_2 , x_3) \sim  {1 \over |x_{23} | } \left(\pa_{x_1^{-}} \pa_{x_2^{-}} -  \pa_{x_1^{-}} \pa_{x_3^{-}} \right) {x^{+}_1 y_{23} + x^{+}_{2} y_{3 1} + x^{+}_3 y_{1 2} \over |x_1 - x_2|^2 |x_1 - x_3|^{2} |x_2 - x_3|}.
}
The total contribution to the pseudo-charge conservation identity is given by the sum of three terms which is zero
\eqn\lastE{
I(x_1 , x_2 , x_3) + I(x_2 , x_3 , x_1) + I(x_3 , x_1 , x_2) = 0.
}
Thus, double trace non-conservation does not contribute to the $\la 0 0 0 \ra$ pseudo-charge
conservation identity.

\appendix{F}{Impossibility to add any further double trace deformations}

Our whole analysis was based on studying the possible terms that appear in the
divergence of the spin four current $J_4$.
We  could wonder whether we can add further double terms to the divergence of
higher spin currents which are not
fixed by the analysis we have already done. In other words, these would be terms
 that appear for higher spin currents but not for the spin four current.
 This would only be possible via the
odd terms, which are the only ones that could have a non-conserved current.
We suspect that all odd terms are fixed by the $J_4$ pseudo-conservation identities, but
we did not prove it. Therefore we will do a separate analysis to argue that we
cannot continuously deform the divergence of the higher spin currents once we have fixed
the $J_4$ one.

Let us imagine that it is possible to introduce an additional parameter for
 the breaking of some  higher spin current. We will focus first on possible double
 trace terms.
Consider the {\it lowest} spin $s$ at which the new term enters\foot{This expression is
schematic. We require and $\epsilon$ tensor in the right hand side if the two operators
$J_{s_1}$ and $J_{s_2}$ have twist one.}
\eqn\newterm{
\nabla.J_s = q  J_{s_1} J_{s_2} + {\rm rest}
}
By assumption $s>4$. The rest denotes terms that are required by $J_4$ non-conservation. This extra term contributes to the  $\la J_s J_{s_1} J_{s_2} \ra_{{\rm odd}}$ three point function. Consider then pseudo-conservation
identity for the $J_4$ current on
 $\la j_{s-2 } j_{s_1} j_{s_2} \ra$. For $s_1, s_2 < s - 2$ we get
\eqn\firstapproxim{
0= q \la j_s j_{s_1} j_{s_2} \ra_{{\rm odd}} + {\rm rest} =   q \la j_s j_{s_1} j_{s_2} \ra_{{\rm odd}}
}
where by ``rest'' we denote the terms that were present when $q=0$.
 These sum up to zero by construction so we have to conclude that $q = 0$.

The argument slightly changes when one of the spins is equal to $s-2$. Below we discuss this case separately for the quasi-boson and quasi-fermion.

\subsec{Case of quasi-fermion}

In this case if, say, $s_1 = s - 2$ we have two possibilities for $s_2$: $2$ or $\tilde 0$. So we write schematically
\eqn\lastchancef{
\nabla .J_s = q (\eps J_{s -2} J_{2} + \pa J_{s-2} \tilde J_0 ) + {\rm rest}
}
where $\epsilon$ is  the three dimension Levi-Civita tensor.
This modifies the correlators
 $\la J_s J_{s-2} J_2\ra_{\rm odd}$ and $\la J_s J_{s-2}  \tilde j_0 \ra_{\rm odd}$.
Consider now pseudo-conservation of the $J_4$ current for
 $\la j_{s - 2} j_{ s - 2 } j_2 \ra$ we get
\eqn\newid{
q \left( \pa_1 \la j_s  j_{s-2} j_2 \ra_{{\rm odd}} + \pa_2 \la j_ {s-2} j_s  j_2 \ra_{{\rm odd}}   \right) = 0.
}
To check this identity it is convenient to introduce in the formula above the dependence on the insertion of $j_2(x)$ and consider the integral
\eqn\strangestep{
\pa_{x_3^{-}} \int d^3 x {1 \over |x - x_3|^4} \left(\pa_1 \la s \ s-2 \ 2(x) \ra_{{\rm odd}} + \pa_2 \la s-2 \ s \ 2 (x) \ra_{{\rm odd}} \right) = 0
}

Using formula \substr\ we get that it is equivalent to the identity
\eqn\simpleident{\eqalign{
\pa_1 \la s \ s-2 \ \pa^{2 \perp }2 \ra_{{\rm fer}} + \pa_2 \la s-2 \ s \ \pa^{2 \perp } 2 \ra_{{\rm fer}}  & \cr
-\pa_1 \la s \ s-2 \ \pa^{2 \perp }2 \ra_{{\rm bos}} - \pa_2 \la s-2 \ s \ \pa^{2 \perp } 2 \ra_{{\rm bos}} &= 0
}}
%
We now take a  light cone limit $\la s_1 \underline{ s_2 \pa^{2 \perp }2 }\ra$.
 More specifically we take the limits
 $\lim_{y \to 0} y |y| \lim_{x^{+} \to 0}$ and  $\lim_{y \to 0} y^2 \lim_{x^{+} \to 0}$, which pick out the boson and fermion pieces respectively.
 Then  one can  check
 that \simpleident\ does not have a solution.

The next case to consider is $\la s-2 \ s-2 \ \tilde 0\ra$. In this case we have
\eqn\newidB{
q \left( \pa_1 \la s \ s-2 \ \tilde 0 \ra_{{\rm odd}} + \pa_2 \la s-2 \ s \ \tilde 0 \ra_{{\rm odd}}   \right) = 0.
}
Again we introduce the explicit dependence on the insertion point $\tilde 0 (x)$ and integrate to get
\eqn\tricksec{
q \int d^3 x {1 \over |x - x_3|^2} \left( \pa_1 \la s \ s-2 \ \tilde 0(x) \ra_{{\rm odd}} + \pa_2 \la s-2 \ s \ \tilde 0(x) \ra_{{\rm odd}}   \right) = 0
}
Using \fermextra\ this becomes
\eqn\simpeq{
q \left[ \pa_1 \la s \ s-2 \ 0 \ra_{{\rm bos}} + \pa_2 \la s-2 \ s \ 0 \ra_{{\rm bos}} \right]  = 0.
}
This identity does not hold and we conclude that $q = 0$.

\subsec{Case of quasi-boson}

In this case if, say, $s_1 = s - 2$ we have two possibilities for $s_2$: $2$ or $0$. So we write schematically
\eqn\lastchanceb{
\nabla.J_s = q (\eps J_{s -2} J_{2} +\eps \pa^2 J_{s-2} J_0 ) + {\rm rest}
}
We  modify the correlators  $\la J_s J_{s-2} J_2 \ra_{\rm odd}$ and $\la J_s J_{s-2} j_0 \ra_{\rm odd}$.
Consider now the $J_4$ pseudo-conservation identity for $\la j_{s - 2} j_{s - 2} j_2 \ra$.   The argument is then identical to the fermion one. We conclude that the first term
in \lastchanceb\ is not possible.
Then we consider the $J_4$ pseudo-conservation on
 $\la j_{s-2} j_{s-2} j_0\ra$. In this case we have
\eqn\newidBb{
q \left( \pa_1 \la j_s  j_{s-2} j_0 \ra_{{\rm odd}} + \pa_2 \la j_{s-2} j_ s j_ 0 \ra_{{\rm odd}}   \right) = 0 .
}
we introduce the explicit dependence on the insertion point $ 0 (x)$ and integrate to get
\eqn\tricksec{
\int d^3 x {1 \over |x - x_3|^4} \left( \pa_1 \la s \ s-2 \ 0(x) \ra_{{\rm odd}} + \pa_2 \la s-2 \ s \ 0(x) \ra_{{\rm odd}}   \right) = 0
}
Using \bosextra\ we find
\eqn\simpeq{
\pa_1 \la s \ s-2 \ \tilde 0 \ra_{{\rm fer}} + \pa_2 \la s-2 \ s \ \tilde 0 \ra_{{\rm fer}} = 0.
}
This identity does not hold and we conclude that $q = 0$.

\subsec{Triple trace deformation}

Here we would like to analyze the new triple trace deformations.
\eqn\lowtriple{
\nabla.J_s = q J_{s_1} J_{s_2} J_{s_3} + {\rm rest}.
}
First notice that this deformation does not affect any of three point functions. Thus, it
does not affect any of the identities that we got using $J_4$. Thus, if our assumption that
the identities for the three point functions fix the odd pieces completely is correct then
any of these terms will not affect any of our conclusions.

Applying arguments similar to the above ones for pseudo-conservation
of $J_4$ on $\la j_{s-2} j_{s_1} j_{s_2} j_{s_3} \ra $ we expect to find that
we cannot add the $q$ deformation. However, we did not perform a complete analysis.

\appendix{G}{Motivational introduction }

In this appendix we give a longer motivational introduction for
the study of the constraints imposed by the higher spin symmetry.

As it is well known,  the structure of theories with massless particles
with spin is highly constrained.
For example, massless particles of spin one lead to the Yang Mills theory,
at leading order in derivatives. Similarly massless particles of spin two
lead to  general relativity. We also  need
   the assumption that the leading order interaction at low energies
is such  that the particles are charged under the gauge symmetries or that
gravitons couple to energy in the usual way.

Now for {\it massive} spin one particles, what
can we say? If we assume that we have a weakly interacting theory for energies
much bigger than the mass of the particle, and we assume a local bulk lagrangian without other
higher spin particles, then we find that the theory should contain at least a
 Yang-Mills field plus a Higgs
particle. We can then view the theory as having  a
spontaneously broken gauge symmetry. Notice
that we are assuming that the theory is weakly coupled.

Now consider a theory that has massless particles with higher spin, $s>2$. If we are
in flat space and the
S-matrix can be defined, then we expect that the Coleman-Mandula theorem should forbid any interaction \refs{\ColemanAD,\HaagQH}\foot{The Coleman-Mandula theorem assumes that we have a finite number
of states below a certain mass shell. For the purpose of this discussion we assume
 that it still applies... }.
Here we are assuming that the couplings of the higher spin fields are such that particles are
charged under the higher spin transformations. This is the case if we have a graviton in the theory, the
same vertex that makes sure that the graviton couples to the energy of the higher spin particle also implies
that the  particles transforms under a higher spin charge.

Now we can consider theories that contain massive particles with spin $s>2$. We assume that we have
a weakly coupled theory at all energies, even at very high energies. In other words,   we assume a suitable
decay of the amplitudes at high energy (such as the one we have in string theory). Then it is likely
that the theory in question is a full string theory.
In other words, we can propose the following conjecture:

{\it A Lorentz invariant theory in three or more dimensions,
 which is weakly coupled\foot{The weak coupling assumption is important.
 In fact, the reader might think of the following apparent counter example. Consider the scattering of higher spin excited states of hydrogen atoms. These are higher spin states, but they are not strings! However, these states
  are not weakly coupled to each other in the particle physics sense.},
  contains a massive particle with spin $s>2$, and  has amplitudes
with suitably bounded behavior at high energies, should be a string theory.}\foot{One probably needs to assume some interactions between the massive
spin particles which makes sure that the particles are ``charged'' under the higher spin symmetry. This would
be automatically true if we also include an $m=0$, $s=2$ graviton. This should also apply to
theories like large $N$ QCD, which have a conserved stress tensor leading to massless graviton in the bulk. }

Of course, there are many string compactifications, so
we do not expect this to fix the theory uniquely. This is precisely the problem
that the old dual models literature tried to solve,
 and string theory was found as a particular
solution. The above conjecture is just that it is the {\it only} solution.
As an analogy, in gauge theories with a spontaneously broken symmetry we can have many realizations of
the Higgs mechanism. However, the paradigm of spontaneously broken gauge symmetry constrains the theories in
an important way. In string theories, or theories with higher spin massive particles, we also expect that
their structure is constrained by the spontaneous breaking of the higher spin symmetry.
For example, we expect that once we have a higher spin particle, we have  infinitely many
 of them. It is notoriously
difficult to study infinite dimensional symmetries. It is even more difficult if there is no unbroken phase
that is easy to study. These ideas were discussed in \refs{\GrossUE,\MooreQE}
 (see \SagnottiQP\ for a recent discussion containing many further references).

Here is where the $AdS$ case appears a bit simpler. In $AdS$ space, as opposed to flat space, it is possible
to have a theory which is interacting (in $AdS$) but that nevertheless realizes the unbroken higher spin
symmetry \refs{\Vasiliev, \Vasilievb, \Vasilievc}.
The fact that these
theories exist can be understood from AdS/CFT, they are duals to free theories \refs{\SundborgWP,\wittentalk,\HolographySun,\KlebanovJA,\SezginPT,\GiombiWH}.
 In any example of
AdS/CFT that has a coupling constant on the boundary we can take the zero coupling limit. In this limit,
 the boundary theory has single trace operators sitting at the twist bound.
  These are all states that are given by  bilinears in the fields. Furthermore,
this forms a closed subsector under the OPE \MikhailovBP .
 Thus we can always find a Vasiliev-type theory as a consistent
truncation of the zero coupling limit of the full theory.
 The Vasiliev-type theories contain only the higher
spin fields (perhaps plus a scalar). They are analogous to the Yang Mills theory without an elementary Higgs particle. For example, if we take the zero $\lambda$ limit of the ${\cal N}=4$ super Yang-Mills theory, we get a certain theory
with higher spin symmetry. Besides the higher spin currents,
 this theory contains lots of other states that
are given by single trace operators which contain more than two field insertions, say four, six, fields in the trace. Here the Vasiliev-like theory is the restriction to the bilinears.

In $AdS_4$ we can realize the dream of constraining the theory from its symmetries.
If the higher spin symmetry is unbroken, we can determine all the correlation functions on
the $AdS$ boundary. This can be done not only at tree level, but for all values of the
bulk coupling constant ($1/N$). This is one way of reading the results of \MaldacenaJN .
Here we are assuming that the $AdS$ theory is such that we can define boundary correlators
obeying the usual axioms of a CFT. In other words, we are assuming the AdS/CFT correspondence.

In Vasiliev-type theories the higher spin symmetry can be broken only by two (or three)
particle
states, which is what we study in this article.

Finally, one would like to study theories that,  besides the higher spin fields,  also
contain enough extra fields that can Higgs the higher spin symmetry at the classical level.
In such theories the mass of the higher spin particles would be nonzero and finite even
for very small bulk coupling.
First one needs to understand what kind of ``matter'' can be added
to the Vasiliev theory and still preserve the higher spin symmetry. Since the higher
spin symmetry is highly constraining,
 it is likely that the only ``matter'' that we can add
is what results from large $N$ gauge theories. As far as we know,
this is an unproven speculation, and it
is closely related to the conjecture above. Note that such gauge theories have a Hagedorn
density of states, so that the bulk theories would look more like ordinary string theories.
Once, this problem is understood, one could consider a case where the Higgs mechanism can
be introduced with a small parameter (as in    ${ \cal  N}=4$ SYM, for example). Here we will have single trace terms in the divergence of the higher spin
currents. This breaking mechanism will be constrained by the higher spin symmetry.
Understanding how it is constrained when the coupling is small will probably give us
clues for how the mechanism works for larger values of the coupling. In addition, it could
give us a way to do perturbation theory for correlators in gauge theories in a
completely on shell fashion.

Whether this idea is feasible or not, it remains to be seen...

\listrefs

\bye